\documentclass[11pt]{article}
\usepackage[]{algorithm2e}
\usepackage{graphicx,wrapfig,lipsum}
\usepackage{graphicx,psfrag,epsf}
\usepackage{amsmath}
\usepackage{enumerate}
\usepackage{amsthm}
\usepackage{bm}
\usepackage{bbm}
\usepackage{titling}
\usepackage{natbib, color}

\usepackage[hyphens]{url}

\usepackage{cprotect}
\usepackage{fancyvrb}
\usepackage{xcolor}

\usepackage{lipsum}

\usepackage{mathtools}
\usepackage{thmtools}
\usepackage{xr}
\usepackage{amssymb}
\usepackage[utf8]{inputenc}
\usepackage[english]{babel}
\usepackage[margin=1in]{geometry}

\newcommand{\blind}{0}

\newenvironment{localsize}[1]
{%
  \clearpage
  \let\orignewcommand\newcommand
  \let\newcommand\renewcommand
  \makeatletter
  \input{bk#1.clo}%
  \makeatother
  \let\newcommand\orignewcommand
}

\SaveVerb{maps}|maps|

\newtheorem{theorem}{Theorem}[section]

\newtheorem{lemma}[theorem]{Lemma}

\renewcommand{\baselinestretch}{1.5}
\newcommand{\distas}[1]{\mathbin{\overset{#1}{\kern\z@\sim}}}%
\title{Computationally Efficient Multivariate Changepoint Detection}
\author{S. O. Tickle, I. A. Eckley, P. Fearnhead}
\date{\today}

\newcommand*{\QEDB}{\hfill\ensuremath{\square}}

\begin{document}

\def\spacingset#1{\renewcommand{\baselinestretch}%
{#1}\small\normalsize} \spacingset{1}

\if0\blind
{
  \title{\bf A computationally efficient, high-dimensional multiple changepoint procedure with application to global terrorism incidence}
  \author{S. O. Tickle$^{1, 3}$\thanks{Email: sam.tickle@bristol.ac.uk, Address: School of Mathematics, Fry Building, University of Bristol, Bristol BS8 1UG, United Kingdom} $^{,4}$ , I. A. Eckley$^2$, P. Fearnhead$^2$
		\hspace{.2cm}\\
    $^1$STOR-i Centre for Doctoral Training, Lancaster University, Lancaster, United Kingdom; \\
    $^2$Department of Mathematics and Statistics, Lancaster University, Lancaster, United Kingdom; \\
    $^3$School of Mathematics, University of Bristol, Bristol, United Kingdom \\
    $^4$Heilbronn Institute for Mathematical Research
}

\maketitle
} \fi

\if1\blind
{
  \bigskip
  \bigskip
  \bigskip
  \begin{center}
    {\LARGE\bf Title}
\end{center}
  \medskip
} \fi

\begin{abstract}
Detecting changepoints in datasets with many variates is a data science challenge of increasing importance. Motivated by the problem of detecting changes in the incidence of terrorism from a global terrorism database, we propose a novel approach to multiple changepoint detection in multivariate time series. Our method, which we call SUBSET, is a model-based approach which uses a penalised likelihood to detect changes for a wide class of parametric settings. We provide theory that guides the choice of penalties to use for SUBSET, and that shows it has high power to detect changes regardless of whether only a few variates or many variates change. Empirical results show that SUBSET out-performs many existing approaches for detecting changes in mean in Gaussian data; additionally, unlike these alternative methods, it can be easily extended to non-Gaussian settings such as are appropriate for modelling counts of terrorist events.
\end{abstract}

\noindent%

%\spacingset{1.45}
\section{Introduction}\label{sec:intro}

The canonical, one-dimensional, changepoint analysis problem has been the focus of substantial research effort for many years. Much initial effort was placed on developing methodology to detect changes in key statistical features, for example changes in mean \citep{Hinkley71}, variance \citep{InclanTiao94}, event rate \citep{Yao86} and distribution \citep{Carlstein88}. More recently, due to the large amount of data now routinely collected, significant focus has been placed on the development of computationally efficient, multiple changepoint search methods. See  \cite{ShiWuRao17}, \cite{EichingerKirch18}, \cite{AnastasiouFryzlewicz19},  \cite{TickleEckleyFearnhead18} and \cite{GrundyKillickMihaylov20} for a selection of recent contributions in this area. %The community has also developed numerous open-source software packages to implement these methods \textcolor{red}{CITES NEEDED HERE}.

In parallel with these developments, there has been a growing adoption of changepoint methods to real world data problems in social and medical settings \cite[e.g.][]{CarrollLawsonZhao19, FarahaniKazemzadeh19, SalmasniaMohabbatiNamdar19}.
%, Kalligerisetal19, Layeghifardetal19, SalmasniaMohabbatiNamdar19, TalaeiKhoeiWilsonKazemi19}. 
Other recent substantive changepoint applications include the maintenance of safe carbon dioxide levels in spacesuits \cite[]{Bekdashetal20}; detecting neuronal activity in calcium imaging data \cite[]{JewellHocking19}; and assessing the effectiveness of interventions to contain the spread of the COVID-19 pandemic \cite[]{Dehningetal20}.

Our work is motivated by analysing the Global Terrorism Database \cite[]{LaFreeDugan07}. This database provides a global historical record of terrorist incidents and is maintained by the National Consortium for the Study of Terrorism and Responses to Terrorism at the University of Maryland. Here, a terrorist event is considered to be  an event `involving ``threatened'' or actual use of illegal force and violence to attain a political, economic, religious or social goal through fear, coercion or intimidation.' We analyse jointly the counts of terrorist events across twelve global regions. 
% building on the Pinkerton Global Intelligence Services (PGIS) database, that have collated all terrorist incidents from 1 January 1970 onwards. PGIS defined terrorism as `events involving ``threatened'' or actual use of illegal force and violence to attain a political, economic, religious or social goal through fear, coercion or intimidation.' 

This application raises two challenges that are common in modern applications. The first is the need to analyse multivariate data, and detect changes that may affect only some of the variates. The second is the need for methods that can detect changes when it is not appropriate to model the data as a change in mean of Gaussian data -- which has been the focus of most multivariate changepoint methods to date \cite[]{SamworthWang18,EnikeevaHarchaoui19,Hahn20}. Simple application of existing methods to these data are unreliable as they detect too many changes; see Appendix B of \cite{Tickle19} for empirical confirmation of this.

The method we develop is based on likelihood ratio tests, and can be applied across a range of modelling scenarios. This makes it stand out from most current competitors, designed to detect change in mean in Gaussian data, and is important for our application where the data are in the form of counts. In such a setting, it is natural to model this using a negative binominal model due to substantial over-dispersion in the data relative to a Poisson model. Whilst not needed for our application, the fact that our method combines likelihood ratio tests for a change on each variate means that it can easily be applied to mixed data settings -- where we may wish to use different models for the different variates which may be of different types (e.g. a mix of continuous, count and categorical data). For the case of a change-of-mean in Gaussian data, we analyse theoretically the properties of our method, and show that it achieves similar asymptotic power to the best possible for this class of models. 

\section{An Introduction to the Global Terrorism Database}\label{sec:gtdintro}

%Terrorism is an increasingly global problem. 
Over the past fifty years, terror activity has seemingly transitioned from being a phenomenon perpetuated by a small number of localised groups to an issue of international significance. One question we may ask is whether there have been specific points or periods of time in which changes in the number of terrorist attacks occur, either regionally or worldwide. Such changes, if confined to certain regions of the globe, may point to specific events which altered the probability of an attack in a particular country or continental area. Alternatively, if a change is seen to affect terrorism worldwide, one might potentially identify events that had a much more wide-ranging scope. Identifying historically impactful events has benefits for future policy-makers, both national and international, particularly in regions of the world where terrorism may be an especially acute issue.

%For a detailed cataloguing of terror events during the last half-century, we turn to 
The Global Terrorism Database \citep{GTD18} is a global historical record of terrorist incidents maintained by the National Consortium for the Study of Terrorism and Responses to Terrorism at the University of Maryland, and the database is copyrighted to the University of Maryland (2018). The database is a compilation of terror events that have occurred worldwide since 1 January 1970. The original platform for the database was the Pinkerton Global Intelligence Services who, from 1970 to 1997, employed a number of researchers to collate and record events from numerous domestic and foreign reports. During this period, an event which came to the attention of the compilers was valid for inclusion if it involved the ``threatened or actual use of illegal force and violence to attain a political, economic, religious or social goal through fear, coercion or intimidation''  \citep{LaFree10}. In addition, any event known to have been perpetrated by a sovereign state was not included.

The Global Terrorism Database has been subject to several analyses in recent years.  \cite{LaFreeDugan07}, \cite{LaFree10} and \cite{LaFreeDuganMiller14} highlighted the higher incidence of terrorism in Europe in the 1970s; a period of unusually high terrorist activity in Latin America between 1980 and 1997; and a more general note regarding the concentration of most incidents within geographic space. Particular events or points in time that may have seen changes in terror activity were also a significant theme of interest. On a global scale, \cite{LaFree10} notes that terrorism tripled between 1976 and 1979, with a doubling just in the final year of this period \citep{LaFreeDugan07}. One of the principle findings of an analysis by \cite{SantifortSandlerBrandt12} %estimate changes in the `arrival rate' of terrorist incidents using a Reversible Jump Markov Chain Monte Carlo approach. One of their principle findings 
suggests that, by 2010, the variation in the mode of terror attacks had declined significantly since 1970, with bombings, typically of non-official, private entities or groups of people, becoming the preferred method of terrorists worldwide. Other analyses of the Global Terrorism Database examine the relationship between terrorism and a wide range of distinct global applications. These include investor sentiment, tourism levels in the United Kingdom, and the worldwide cost of debt; see \cite{Drakos10}, \cite{Mao19} and \cite{ProcaskyUjah16}, among others. %Meanwhile, other analyses of similar data include \cite{ClausetYoung05}, who focus on the severity, rather than number, of incidents for a given area at a given moment in time.

The database has also been used to analyse terrorism at a much more local level. For example, \cite{LaFree10} discusses the activities of the Armenian Secret Army for the Liberation of Armenia (ASALA). %In 1983, following a period of specifically targeting supporters and perceived supporters of the government in Turkey, and largely avoiding non-Turkish citizens, an attack by ASALA at an airport in Paris went awry when a bomb detonated early. This led to a substantially less discriminate targeting strategy from the group, which subsequently lost legitimacy in the eyes of its supporters. 
The link between government policy change and a change in the probability of a terrorist attack has also been examined. For example, \cite{LaFree10} used the database to perform an analysis on six policies enacted by the British government to attempt to reduce terror activity in Northern Ireland. It was discovered that three of these policies in fact resulted in a significant increase in terrorism, with only one leading to a reduction.
%%Aside from the principal investigators, other authors have also canvassed the question of change in terror activity using the Global Terrorism Database. 
 Meanwhile, \cite{RaghavenGalstyanTartakovsky13} focus on the activities of the Revolutionary Armed Forces of Colombia - People's Army (FARC) between 1998 and 2006. They conclude that, from the end of June 1998, funding from the United States to combat the drug economy in Colombia had an impact in reducing FARC's activities in the short to medium term.

\begin{figure}[h]
\begin{minipage}[t]{\textwidth}
\vspace{-10pt}
\begin{center}
\includegraphics[width=0.98\linewidth,keepaspectratio=true]{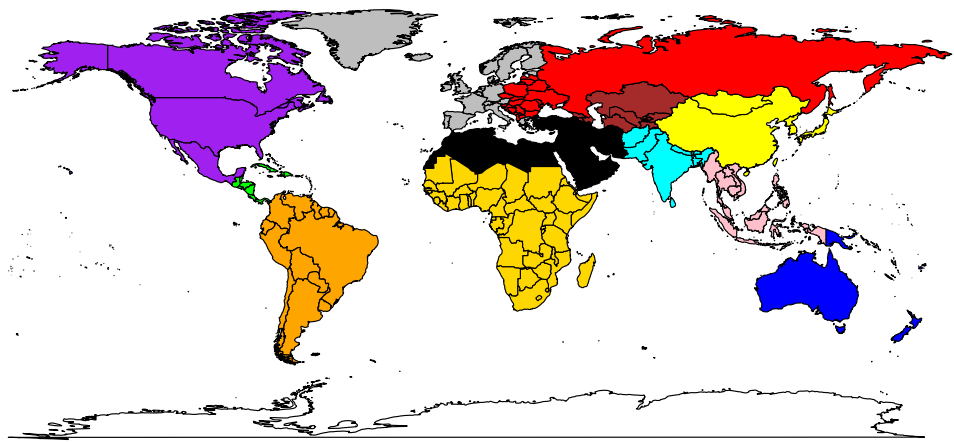}
\end{center}
\vspace{-90pt}
\caption{Definition of twelve regions used when analysing the global terrorism data set -- each region is denoted by a different colour. This map was created with the aid of the \protect\UseVerb{maps} package of \cite{BeckerWilks18}.}
\label{worldmap}
\end{minipage}
\end{figure}

We use the Global Terrorism Database to provide event incidence in twelve regions: Australasia \& Oceania, Central America \& Caribbean, Central Asia, East Asia, Eastern Europe, Middle East \& North Africa, North America, South America, South Asia, Southeast Asia, Sub-Saharan Africa and Western Europe. Given that these political terms may be somewhat fluid geographically, we show this division pictorially in Figure~\ref{worldmap}. Note that these regions are as defined by the compilers of the database. It would also be possible to carry out our later analysis of the database on a much finer scale, for example, by country. However, given that many sovereign states have undergone substantial border changes in the last fifty years, such an analysis would likely require a great deal of care.

\begin{figure}[h]
\begin{minipage}[t]{\textwidth}
\begin{center}
\includegraphics[width=0.98\linewidth,keepaspectratio=true]{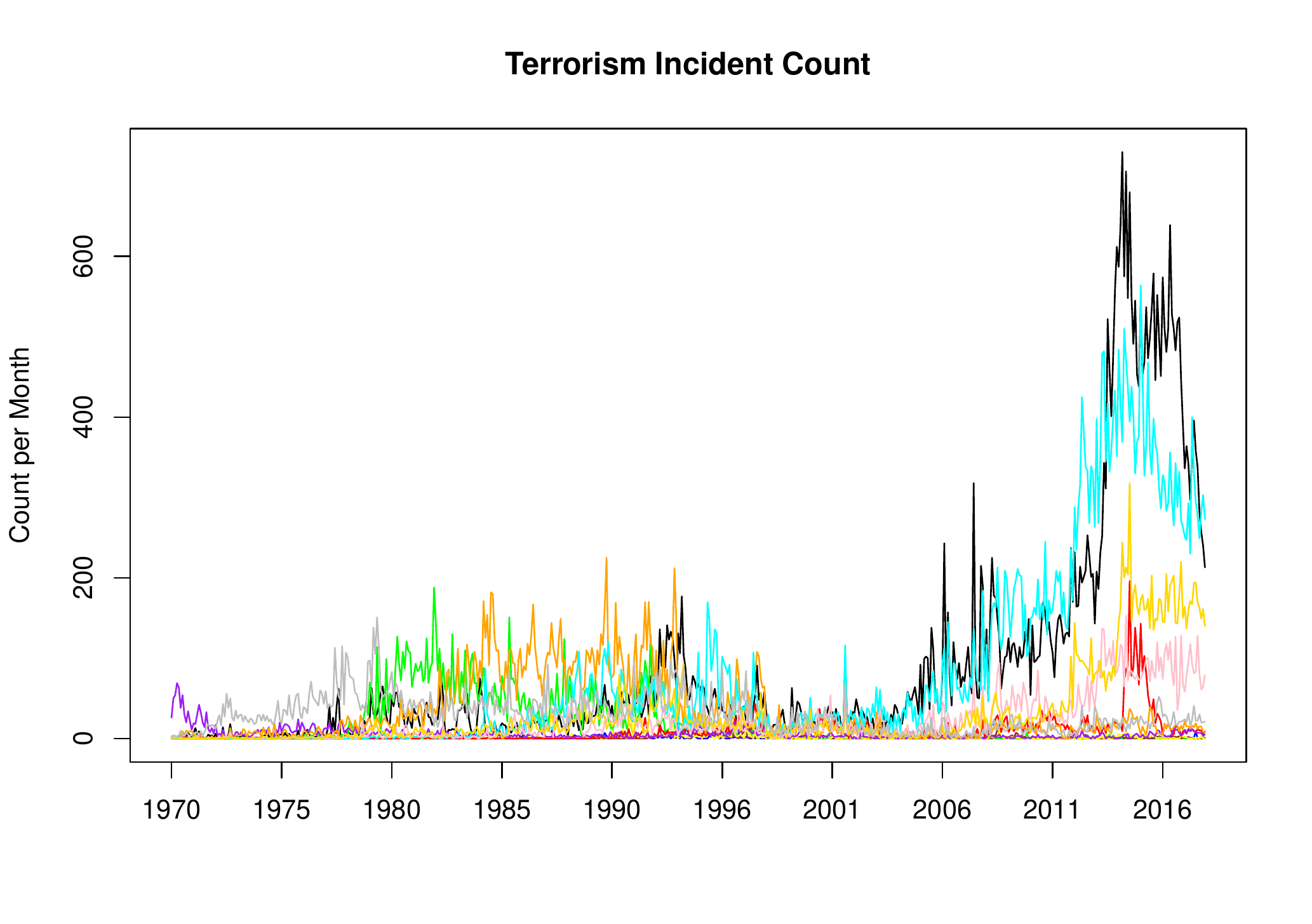}
\end{center}
\vspace{-30pt}
\caption{Terrorism incident count per month for each of the 12 regions in Figure~\ref{worldmap}. Note that the series' colours match those of Figure~\ref{worldmap}: Australasia and Oceania (blue); Central America and Caribbean (green); Central Asia (brown); East Asia (yellow); Eastern Europe (red); Middle East and North Africa (black); North America (purple); South America (orange); South Asia (cyan); Southeast Asia (pink); Sub-Saharan Africa (gold); and Western Europe (grey).}
\label{incidentcount}
\end{minipage}
\end{figure}

For each of the twelve regions, we aggregate all incidents for each month to produce one univariate time series of counts for each region. Each of these is of length 564, one for each month between January 1970 and December 2017 inclusive. Note that 1993 is not included, as that year's data are missing from the publicly-available copy of the database. The resulting incident count by region is shown in Figure~\ref{incidentcount}.

We can see from Figure~\ref{incidentcount} that there appear to be periods of time in which terrorism abruptly increases or decreases in specific regions. Whilst the question of locating changepoints in the database could be answered using univariate changepoint methods, we seek to analyse these series jointly and thereby potentially identify subsets of series within which changes occur. Such a setting is inherently more computationally complex, due to the rapid increase in possible changepoint combinations that can occur in a high-dimensional setting. This is the challenge we seek to address, proposing a novel computationally efficient approach to identify changepoints in multivariate data sequences. 

\section{Background}\label{sec:background}

%We begin by briefly reviewing some of the existing literature on multivariate changepoint analysis. 
Whilst detecting changes in univariate data sequences has a long history, there has been much less work on methods for detecting potentially multiple changepoints in multivariate datasets. Univariate approaches can be easily adapted to the multivariate setting if we are willing to assume all variates change at each changepoint \cite[e.g.][]{Wessman98}.  However, this may not be appropriate in applications where some, but not all, variates are affected by each changepoint; or where it is not known \emph{a priori} whether a change will only affect a very small number or many of the variates.

Within the multivariate changepoint setting, the change in mean problem has to date received the most substantial focus. In this setting, evidence for a change in a single series can, for example, be quantified using CUSUM statistics -- a weighted difference in the empirical mean before and after the potential changepoint. The simplest ways of combining evidence across time series are to (i) perform some form of averaging of the CUSUM statistics; or (ii) take the maximum value of the CUSUM statistics.  Naturally, the detection boundaries, i.e.\ how large a change in mean is needed before the presence or absence of change can be determined with probability tending to 1, are very different for approaches (i) and (ii), with the first being able to detect small changes that affect most variates, and the latter requiring at least one variate to change by a sizeable amount.  \cite{EnikeevaHarchaoui19} investigate the detection boundary for a change in mean in a high-dimensional asymptotic setting where the number of variates, $d$, the number of variates that change and the number of observations per variate increase. They show that there are two regimes depending on whether the number of variates that change increases faster than $\sqrt{d}$, or at or slower than a $\sqrt{d}$ rate. We call the former regime the dense change regime, and the latter the sparse change regime. Informally, approach (i) works well in the dense regime, while (ii) works well in the sparse regime.

%study the properties of both approaches under an asymptotic regime where both the number of variates and the number of observations per variate increase; indeed, in the case of the maximum, they consider a more general framework, namely the maximum across all subsets of variates (which they label as a scan statistic). In particular, which of the methods is better depends on the proportion of variates that undergo a sizeable change. %If we let $d$ be the number of variates, and we say a change is sparse if it affects $o(d^{1/2})$ of the variates, and dense otherwise, then methods based on averaging CUSUM statistics are able to detect smaller changes in the dense setting. By contrast, using the maximum can detect smaller changes in the sparse setting. \cite{EnikeevaHarchaoui19} propose combining these two approaches in order to have a high detection chance across all types of change. 

In recent years, a number of other approaches have been proposed that seek to strike a balance between the (i) mean and (ii) max CUSUM-based approaches. For example, \cite{ChoFryzlewicz15} and \cite{Cho16} sum only CUSUM statistics that exceed a certain threshold; meanwhile, \cite{SamworthWang18} consider sparse projections of the data, which is equivalent to using a weighted average of CUSUM statistics. These approaches can demonstrate strong empirical performance, but neither has been shown theoretically to simultaneously work as well in both the dense and sparse settings. For example, the method of \cite{SamworthWang18} was designed for detecting sparse changes, and its theory establishes strong performance in precisely that setting. It is the development of an approach that seeks to work well in both settings that we introduce below. The one current method that has been shown to work well in both dense and sparse settings is that of \cite{EnikeevaHarchaoui19} which combines a test statistic based on combining all CUSUM statistics with one based on the largest $p$ statistics and scans over all choices of $p$.

To this end, suppose that the data sequence for each variate, $\left(y_{i,j}\right)_{j=1}^n$ for $i = 1, \ldots, d$, within the dataset, $\textbf{y}_{1:n}$, can be segmented by changepoints, which are often shared across variates within the data. So for the global terrorism data $y_{i,j}$ corresponds to the count of terrorist events in region $i$ and time point $j$, and we have $d=12$ and $n=564$. % Following ideas introduced by \cite{Pickering16}, 
We define the set of changepoints to be points where at least one variate undergoes a change. Therefore, for each changepoint there is an associated \textit{affected set} of variates which undergo a change.

Formally, let $0 = \tau_0 < \tau_1 < \ldots < \tau_{m} < \tau_{m+1} = n$ be the changepoints, with corresponding affected sets $\mathcal{S}_1, \ldots, \mathcal{S}_{m}$. Note that it is possible for a given variate to be affected at more than one changepoint, so these affected sets are allowed to overlap. We will assume a parametric model for the data within a segment for each variate, and further that the segment parameter for this model only changes at changepoints which affect that variate. To simplify the exposition, assume that the data are conditionally independent given the segment parameters. In other words, we have
\begin{equation}\label{genproc}
y_{i,j} \sim g(.|\mu_{i,k}),
\end{equation}
for some family of densities $g(.|.)$, where $k = \left|\left\{v : \tau_v < j \right\}\right|$. Here, $k$ denotes the segment associated with time-point $j$, and $\mu_{i,k}$ is the associated segment parameter for series $i$. We have $\mu_{i,k}=\mu_{i,k+1}$ unless $i \in \mathcal{S}_k$: the $k^{\mbox{th}}$ and $(k+1)^{\mbox{th}}$ segment parameters of series $i$ are identical unless the $k^{\mbox{th}}$ change affects series $i$.

Our definition of the within-segment data-generating processes given above allows our method to solve for a wide class of possible changepoint problems. For instance, the well-studied canonical problem of Gaussian change in mean can be included by setting 
\begin{align*}
y_{i,j} \sim N\left(\mu_{i,k}, \sigma_i^2\right),
\end{align*}
where $\mu_{i,k}$ is the mean of the signal in the $i^{th}$ variate following the $(k-1)^{th}$ changepoint which affects variate $i$, and $\sigma_i^2$ is some (known) variance. However, for the Global Terrorism Database example discussed in Section~\ref{sec:gtdintro}, this classical problem setup is inappropriate. Given that this particular application is comprised of count data across multiple regions, we can use a negative binomial likelihood such that
\begin{align*}
\mathbb{P}(y_{i,j} = y) = {y + r_i - 1 \choose y} \times (1 - p_{i,k})^y p_{i,k}^{r_i},
\end{align*}
which we can refer to as 
\begin{align*}
y_{i,j} \sim \mbox{Neg-Bin}(r_i, p_{i,k}), \mbox{ for } \tau_{k-1} + 1 \leq j \leq \tau_k.
\end{align*}
Here, $r_i$ governs the amount of {\em over-dispersion}, relative to a model, in the $i^{th}$ variate. For a fixed $r_i$, $p_{i,k}$ then determines the mean of the $i^{th}$ variate following the $(k-1)^{th}$ changepoint. %Note that the nomenclature for this parameter arises from the classical consideration of the negative binomial, which models the number of failures in a sequence of independent Bernoulli trials, each with success probability $p$, until the $r^{th}$ failure is observed. 
%In our later analysis of the Global Terrorism Database, we assume such a negative binomial likelihood. % with fixed over-dispersion parameter but changing mean. %$p_{i,k}$, which in this instance can be thought of as proportional to the probability of a terrorist attack in each region.

%We remark that, in the Global Terrorism Database example, the number of incidents in each region are not conditionally independent of one another. However, as discussed in \cite{Tvetenetal20}, for correlations between series less than 0.5, there is little impact on power or estimation accuracy in the multivariate changepoint detection setting. As we demonstrate in Section~4 of the Supplementary Material, the correlations between the residuals of different series in our real data example are all less than 0.5.

Whilst our aim is to jointly detect the number and location of all changepoints, as well as the variates that are affected at each change, we  first introduce and develop theory associated with the analysis of our approach under the assumption that there is at most one changepoint. This will subsequently be extended within a binary segmentation algorithm to detect multiple changepoints.

\section{SUBSET}\label{sec:themethod}

\subsection{Detecting a Single Changepoint} \label{ssec:amoc}

We begin with a derivation of the test statistic used by SUBSET in the single change setting. The log-likelihood ratio statistic for detecting a changepoint at time $\tau$, affecting variates in set $\mathcal{S}$, is
\small \[
R(\tau,\mathcal{S})=2\left[ \sum_{i \in \mathcal{S}}
\left\{ \max_{\mu} \sum_{t=1}^{\tau} \log g(y_{i,t}|\mu)
+\max_{\mu} \sum_{t=\tau+1}^{n} \log g(y_{i,t}|\mu)
-\max_{\mu} \sum_{t=1}^n \log g(y_{i,t}|\mu)
\right\}
\right].
\] \normalsize
To simplify the notation, let $\mathcal{C}(y_{i,s:t})=-2 \max_{\mu} \sum_{j=s}^{t} \log g(y_{i,j}|\mu)$, and 
\[
D_{i,t}=\mathcal{C}(y_{i,1:n})-\mathcal{C}(y_{i,1:t})-\mathcal{C}(y_{i,t+1:n})
\]
denote the contribution to the log-likelihood ratio statistic from series $i$ if it is assumed to change at time $t$. Then 
\[
R(\tau,\mathcal{S} )= \sum_{i \in \mathcal{S}} D_{i,\tau}.
\]
Directly using the log-likelihood test statistic is complicated, due to the fact we do not know $\tau$ or $\mathcal{S}$. In addition, different choices for $\mathcal{S}$ will allow for different numbers of variates to change. We therefore consider a penalised version of the test statistic, where the penalty depends on the number of variates that change, $|\mathcal{S}|$. We then maximise over possible choices of $\tau$ and $\mathcal{S}$. That is, we use $\max_t S_t$ as our test statistic where, for $t=1, \ldots,n-1$,
\[
S_t=\max_{\mathcal{S}} \left\{ \sum_{i \in \mathcal{S}} D_{i,t} - \mbox{Pen}(|\mathcal{S}|) \right\},
\]
for some suitable penalty function $\mbox{Pen}(\cdot)$. 

For both theoretical and computational reasons (see Section~\ref{ssec:theory} and Section~\ref{sec:postproc} of the Supplementary Material respectively), we suggest a piecewise linear penalty of the form
\[
\mbox{Pen}(p)=\min\{ \beta + \alpha p, K\},
\]
for some suitable constants $\alpha$, $\beta$ and $K$. We then detect a change if $\max_t S_t>0$, with the location at $\hat{\tau} = \arg \max S_t$ and the set of estimated affected variates is given by 
\[
\arg \max_{\mathcal{S}}\left\{ \sum_{i \in \mathcal{S}} D_{i,\hat{\tau}} - \mbox{Pen}(|\mathcal{S}|)\right\}.
\]

We choose a piecewise linear penalty as this makes the maximisation over $\mathcal{S}$ computationally efficient. In particular, we can define $D_{i,t}^{'}=\max\{D_{i,t}-\alpha,0\}$, and then 
\[
S_t=\max\left\{\sum_{i=1}^d D_{i,t}^{'}-\beta, \sum_{i=1}^d D_{i,t}-K\right\}.
\]
The two terms in the maximisation above correspond to the two different linear regimes in the penalty function. As we shall see the $\beta+ \alpha p$ component of the penalty function, previously considered by \cite{Pickering16} as a means of penalising changes across both time and variates, determines the test statistic's behaviour for detecting sparse changes. Meanwhile, the constant term, $K$, is needed to improve power for detecting dense changes. If $\xi = \arg \max_{t} \sum_{i = 1}^d D_{i,t}^{'} - \beta$ and $S_{\xi} > 0$, then we say that we have detected a \textit{sparse} change, with evidence for a change only in those variates $i$ such that $D_{i,\xi}^{'} > 0$. If, however, $\eta = \arg \max_{t} \sum_{i=1}^d D_{i,t} - K$ and $S_{\eta} > 0$, then this is identified as a dense change and all variates are labelled as affected. 

%$S_t = \sum_{i = 1}^d D_{i,t}^{'} - \beta > 0$, then we say that we have detected a \textit{sparse} change, with evidence for a change only in those variates $i$ such that $D_{i,t}^{'} > \alpha$. If, however, $S_t = \sum_{i=1}^d D_{i,t} - K > 0$, then all changes are labelled as affected by the estimated changepoint. In this situation, the change is described as \textit{dense}.

\subsection{Theory for a Change in Mean}\label{ssec:theory}

To understand the behaviour of the test statistic for a single change, and obtain guidelines for choosing the constants that define our penalty function, we study its theoretical properties for the canonical change in mean problem with Gaussian noise and a common, known variance, $\sigma^2$. This means that $D_{i,t}$ is $\chi^2_1$-distributed 
%chi-squared distributed with 1 degree of freedom% 
when no changepoint is present. The results we present are also useful for choosing the constants of the penalty function in cases where $D_{i,t}$ is based on a likelihood ratio test for the change of a single parameter, when $D_{i,t}$ would be approximately $\chi^2_1$-distributed  %chi-squared distributed with 1 degree of freedom 
if there is no change.

As we are considering just a single change, we will simplify notation so that $\mu_{i,1}$ is the initial mean of series $i$. If there is a change, $\mu_{i,2}$ will be the mean after the change, and $\mu_{i,1} = \mu_{i,2}$ if $i \notin \mathcal{S}_1$. Thus, the data-generating model is
\begin{equation}\label{amocsetup}
Y_{i,j} = \epsilon_{i,j} + \begin{cases}\mu_{i,1} \mbox{ for } 1 \leq j \leq \tau, \\ \mu_{i,2} \mbox{ for } \tau + 1 \leq j \leq n, \end{cases} \mbox{ for } i \in \left\{1, \ldots, d \right\}
\end{equation}
where the $\epsilon_{i,j}$, for $i = 1, \ldots, d$, and $j = 1, \ldots, n$ are a set of centered, independent and identically distributed Gaussian random variables.

For this particular problem, we have that 
\[
\mathcal{C}\left(y_{i,s:t}\right) = \frac{1}{\sigma^2}\sum_{j=s}^t \left(y_{i,j} - \frac{1}{t - s + 1} \sum_{k = s}^t y_{i,k}\right)^2, 
\]
such that
\[
D_{i,t} = \frac{1}{\sigma^2} \left[\frac{1}{t} \left(\sum_{k=1}^t y_{i,k}\right)^2 + \frac{1}{n-t} \left(\sum_{k=t+1}^n y_{i,k}\right)^2 - \frac{1}{n} \left(\sum_{k=1}^n y_{i,k}\right)^2\right].
\] 
 We use these expressions to establish false positive and detection probability results in the single change setting under Gaussian noise when $\max_t S_t$ is taken as the test statistic. 

Our first theoretical contribution concerns the false positive rate of the chosen test statistic.

\begin{theorem}\label{allparameterchoices}
Suppose we are in setting (\ref{amocsetup}), and that in addition $\mu_{i,1} = \mu_{i,2}$ $\forall i$ and $Var\left(\epsilon_{i,j}\right) = 1$ $\forall i,j$. Take $\alpha = 2 \log d$, $\beta = \left(J + \epsilon\right) \log n$ and $K = \beta + d + \sqrt{2 \beta d}$ for some $\epsilon > 0$; then
\[
\mathbb{P}\left(\max_t S_t > 0\right) \leq C n^{1 - \frac{J}{2} -\epsilon/2},
\]
where $C$ is an absolute constant bounded above for all $d > 1$.
\end{theorem}
\noindent \textbf{Proof}: See Section~\ref{sec:proofs} of the Supplementary Material.

This result therefore provides guidance on suitable choices of the penalty for the Gaussian setting. We note that if $J \geq 2$ this will provide a test whose false positive rate will tend to 0 as $n\rightarrow \infty$. However the bound on the false positive rate ignores the strong positive correlation in $S_t$ for different values of $t$, and thus will be conservative. As such we suggest using simulation to choose $\beta$, and the corresponding value of $K$. This can be done for a given value of $n$ and $d$ by simulating multiple datasets with no change, and setting $\beta$ to, e.g., the value which gives no detected changes for $95\%$ of simulated datasets.
%
%This result suggests setting $\alpha=2\log d$ and gives a relation between the choice of $K$ and $\beta$, in that the choice of $K$ is a function of $\beta$. As the proof highlights, this relationship is based on making the false positive rate for dense and sparse changes comparable. Taking $J\geq 2$ in the definition of $\beta$ then gives a test whose false positive rate will tend to 0 as $n\rightarrow \infty$. However the bound on the false error rate ignores the strong positive correlation in $S_t$ for different values of $t$, and thus will be conservative. As such we suggest using simulation to choose $\beta$, and the corresponding value of $K$. This can be done for a given value of $n$ and $d$ by simulating multiple datasets with no change, and setting $\beta$ to e.g. the value which gives no detected changes for $95\%$ of simulated datasets. 
%
We note also that Theorem~\ref{allparameterchoices}, under a scenario of no change, uses the marginal $\chi^2$-distribution of $D_{i,t}$. This suggests that setting $\alpha=2 \log d$ and using a simulation approach to choose $\beta$ (and hence $K$) would work similarly in other settings when the marginal distribution of $D_{i,t}$ is approximately chi-squared with one degree of freedom. This would be possible, for example, in the common scenario of detecting a change in a single parameter using a likelihood-ratio test.
Given these penalty choices, we can next state a result on the power of this procedure.

\begin{theorem}\label{sufficiencycondition}
Assume that we are again in setting (\ref{amocsetup}) with $\sigma^2 = 1$, and now we have that $\mu_{i,1} \neq \mu_{i,2}$ whenever $i \in \mathcal{S}_{1} \subseteq \{1, \ldots, d\}$. Let $\Delta_i := \left|\mu_{i,2} - \mu_{i,1} \right|$. Then for $2> \delta > 0$ and $a = \max\{n, d\}$, we have that $\mathbb{P}\left(\max_t S_t > 0\right) \geq 1 - (a)^{-\delta}$, providing that, for $K_{\mathcal{S}_{1}} := \beta + |\mathcal{S}_{1}|\alpha$
\[
n\theta\left(1 - \theta\right) \sum_{i \in \mathcal{S}_{1}} \left(\Delta_i\right)^2 \geq \min\{V_S,V_D\},
%\begin{cases} V_S &\text{ for a sparse change} \\  V_D &\text{ for a dense change}, \end{cases}
\]
where $V_S := 4 \delta \log a + K_{\mathcal{S}_{1}} - |\mathcal{S}_{1}| + 2 \sqrt{\delta \log a \left(4 \delta \log a + 2K_{\mathcal{S}_{1}} - |\mathcal{S}_{1}| \right)}$, $V_D := 4 \delta \log a + K - d + 2\sqrt{\delta \log a \left(4 \delta \log a +2 K - d\right)}$ and $\theta = \frac{\tau}{n}$. %Additionally, $2 > \delta > 0$ is required in the dense setting. 
\end{theorem}
\noindent \textbf{Proof}: See Section~\ref{sec:proofs} of the Supplementary Material.

The two constants, $V_S$ and $V_D$, correspond to the ability of our method to detect either a sparse change or a dense change; and come from considering the event that our test statistic is positive for a penalty that is just $\alpha |\mathcal{S}_1|+\beta$ or just $K$ respectively. It is by using a penalty that is the minimum of these two that gives good performance across both sparse and dense change setting. To see this, and to help understand the result, consider the behaviour of $V_S$ and $V_D$ in an asymptotic setting where $n$ and $d$ increase, with $d$ increasing at a polynomial rate in $n$, and the number of variates that change increasing at a polynomial rate in $d$. In this case we have $V_S/(2|\mathcal{S}_1|\log d)\rightarrow 1$ and $V_D/\sqrt{d \log n}\rightarrow c$ for some constant $c$. Thus we have adaption to sparse settings, where $V_S$ will be smaller if $|\mathcal{S}_1|=o(d^\gamma)$ for some $\gamma<1/2$, and dense settings where $V_D$ is smaller if $|\mathcal{S}_1|$ is $O(\sqrt{d})$ or larger.

\subsection{Relationship to Other Multivariate Changepoint Tests}\label{ssec:othermeths}

For the change in mean setting, it is possible to draw strong comparisons between our approach and other multivariate changepoint tests, the main difference being in terms of how the methods aggregate evidence for a change across different variates. These alternative approaches use the CUSUM statistic for each variate within the dataset, defined, in the known $\sigma^2$ case, as
\[
W_{i,t} = \frac{1}{\sigma}\sqrt{\frac{t \left(n - t\right)}{n}} \left|\frac{1}{n-t} \left(\sum_{j = t + 1}^n y _{i,j}\right) - \frac{1}{t} \left(\sum_{j=1}^t y_{i,j}\right) \right|,
\]
for $i = 1, \ldots, d$ and $t = 1, \ldots, n - 1$. Note in particular that $D_{i,t} = W_{i,t}^2$. Therefore, for the Gaussian change in mean setting, our test statistic can be expressed in terms of the CUSUM statistic as
\[
S_t = \max\left\{\sum_{i=1}^d \max\{W_{i,t}^2 - \alpha, 0\} - \beta, \sum_{i=1}^d W_{i,t}^2 - K \right\}.
\]
For comparison, three previously proposed test statistics, which we refer to herein as \textbf{Mean} \citep{Groen13}, \textbf{Max} \citep{Groen13} and \textbf{Bin-Weight} \citep{ChoFryzlewicz15} are, respectively
\[
S_t^{(\mbox{mean})} = \frac{1}{d} \sum_{i = 1}^d W_{i,t} - \beta, ~~
S_t^{(\mbox{max})} = \max_i W_{i,t} - \beta, ~~ %\mbox{ and }
S_t^{(\mbox{bin-weight})} = \sum_{i=1}^d W_{i,t} \mathbbm{1}\left\{W_{i,t} > \alpha \right\} - \beta.
\]
From the results in \cite{EnikeevaHarchaoui19}, we know that $S_t^{(\mbox{mean})}$ will have high power for dense changes which affect most series, but lose power for sparse changes where few variates change. By comparison, $S_t^{(\mbox{max})}$ will have higher power in the sparse case and lower power in the dense case. These can be combined to produce a test with high power across both cases \cite[see][though their proposed method combines different test statistics to mean and max cusum]{EnikeevaHarchaoui19}. %Meanwhile, empirically, the behaviour of $S_t^{(\mbox{bin-weight})}$ depends on the choice of $\alpha$. If $\alpha = \mathcal{O}\left(\sqrt{d}\right)$ then it has high power for sparse changes, whereas if $\alpha$ is fixed as we increase $d$ it will have high power for dense changes.
The Bin-Weight method is closest to the one that we propose, particularly if we set $\alpha=\sqrt{2\log d}$, which is equivalent to the threshold we use. Other than using CUSUM statistics rather than their squares, there are two main differences between Bin-Weight and our method. The first is that as $W_{i,t}$ increases its contribution to $S_t$ will jump from 0 to $W_{i,t}$ when it first exceeds the threshold $\alpha$; by comparison our approach uses a soft-threshold of $W_{i,t}^2$, which avoids such a jump and thus reduces the variability of the test statistic. Second, for a choice of $\alpha=\sqrt{2\log d}$, Bin-Weight will lose power in dense scenarios compared to our approach which caps the overall penalty at $K$. 

\subsection{Sparse and Ubiquitous Binary Segmentation in Efficient Time}\label{ssec:SUBSET}

We here formally introduce SUBSET (\textbf{S}parse and \textbf{U}biquitous \textbf{B}inary \textbf{S}egmentation in \textbf{E}fficient \textbf{T}ime), the full procedure for the use of the test statistic $\max_t S_t$ given in Section~\ref{ssec:amoc}. Given this thresholded penalty approach, SUBSET is designed to detect both sparse and dense changes, the latter of which are labelled by SUBSET as affecting all variates within the data.

In order to detect multiple changes within the data, SUBSET embeds the test for detecting a single changepoint within Wild Binary Segmentation \citep{Fryzlewicz14}. When implementing this procedure, we recommend setting $\alpha=2\log d$, keeping the same relationship between $\beta$ and $K$, and then tuning $\beta$ so that the wild binary segmentation procedure has an appropriate false positive rate for a given $n$ and $d$ on data simulated with no change.
%namely a very close variant of Wild Binary Segmentation 2 \citep{Fryzlewicz19}. This enables a fast, approximate search for changes in the multivariate setting. We randomly generate $M$ intervals of the dataset on which we then subsequently compute the test statistic and search for the most significant point in the dataset across all intervals. We label the resulting point as a change if the penalty in the sparse or dense setting is exceeded. The procedure then repeats either side of the change. For a sensible choice of $M$ - for example, we use $M = 5$ in the simulations in Section~\ref{sec:sims} and $M = 10$ for the real data example in Section~\ref{sec:real} - this results in a computationally efficient procedure with an execution time which is linear in the number of entries in the dataset. (See Section~\ref{sec:simstudyadditional} for empirical confirmation of this.)

An issue with SUBSET is that while the estimates of $\boldsymbol{\hat{\tau}}$ tend to be fairly reasonable, when estimating multiple changes the estimates of $\boldsymbol{\hat{\mathcal{S}}}$ are prone to error due to masking from other changepoints. This is especially true for variates for which there may be a particularly strong change at a nearby time point. For sparse changes, we propose using a post-processing step where we individually analyse data from each variate conditional on the set of estimated changes, $\boldsymbol{\hat{\tau}} = \left(\hat{\tau}_1, \ldots, \hat{\tau}_{\hat{m}}\right)$. When analysing a single variate, we only allow changes to occur within the set $\boldsymbol{\hat{\tau}}$. We detect the changes by minimising the univariate version of our penalised cost, that is introducing a change in a given variate if it reduces the cost by at least $\alpha$. Formally, for variate $i$, we find
%That is, when a point is returned as being the most significant by the single change detection method (when in fact more than one change is present), other changes present in the stream can lead to the estimated affected set being corrupted by a particularly strong change in a given variate at another changepoint.
%Therefore, a post-processing step is required. Given a series of changepoints $\hat{\tau}_1, \ldots, \hat{\tau}_{\hat{m}}$, for some estimated number of changes $\hat{m}$, and corresponding estimated variate sets $\hat{\mathcal{S}}_1, \ldots, \hat{\mathcal{S}}_{\hat{m}}$, we can run a univariate, multiple change procedure through each variate to determine which changes are truly significant in that variate. For example, we may use a dynamic programming approach with a constrained objective (i.e. we can only fit changes for each variate within the set $\{\hat{\tau}_1, \ldots, \hat{\tau}_{\hat{m}}\}$) where a penalty of $\beta$ is added at each changepoint; see, for instance, \cite{TickleEckleyFearnhead18} for details on how this works in practice. 
\[
\arg \min_{0 \leq m^{'} \leq \hat{m}; \{\xi_1, \ldots, \xi_{m^{'}}\} \subseteq \boldsymbol{\hat{\tau}}} \sum_{k=1}^{m^{'}+1} \left[\mathcal{C}\left(y_{i,(\xi_{k-1}+1):\xi_k}\right) + \alpha \right].
\]
This can be done efficiently using dynamic programming. See, for example, Section 2 of \cite{TickleEckleyFearnhead18} for details.

Pseudo-code for the full SUBSET algorithm is provided in Section~\ref{sec:postproc} of the Supplementary Material.

%For the specific post-processing step we use here to complete the SUBSET procedure, please see Section~\ref{sec:postproc}. We include the post-processing step in the implementation of SUBSET in Sections~\ref{sec:sims} and~\ref{sec:real}, however for brevity we detail the SUBSET procedure without the post-processing step in Algorithm 3.

%We remark that under this procedure it may be the case that we check for the possibility of a single changepoint within an arbitrary region of the dataset which contains no true change. In this case a slight modification to Theorem~\ref{allparameterchoices} is required.

%\begin{corollary}\label{multitypeI}
%Consider the setting of Theorem~\ref{allparameterchoices}. Using the SUBSET procedure with the penalties $\beta, \alpha$ and $K$ as derived in Theorem~\ref{allparameterchoices}, with $J=4$, gives that the probability of erroneously placing an estimated changepoint within the dataset is bounded above by $C n^{-\epsilon/2}$, where $C$ is an absolute constant bounded above for all $d > 1$.
%\end{corollary}
%\noindent \textbf{Proof}: See Section~\ref{sec:proofs}.

\section{Simulations}\label{sec:sims}

We examine the properties of the SUBSET method against the CUSUM aggregation procedures discussed in Section~\ref{ssec:othermeths}. In addition, we compare these methods against \textbf{Inspect} \citep{SamworthWang18}, a recent leading approach for detecting changes in mean under Gaussian noise for high-dimensional series. To implement Inspect, we use code from the \verb|InspectChangepoint| package \citep{WangSamworth16}.

All simulations were run in R using a Linux OS on a 2.3GHz Intel Xeon CPU. We examine multivariate series with pairwise independent Gaussian noise with variance 1 and count data generated according to a negative binomial likelihood model under various different dispersion parameters. For all scenarios considered, 200 repetitions were simulated.  The threshold penalty for Inspect and the $\beta$ values for SUBSET and the CUSUM-based methods were computed using simulations from the null model, such that the false alarm rate was fixed at 5\%. For comparability with SUBSET, the $\alpha$ value for Bin-Weight was taken to be $\sqrt{2 \log n}$. For further justification of this choice, see \cite{SamworthWang18}. %We remark that, given the distinction between $\alpha$ and $\beta$ in the setup of Bin-Weight, as with SUBSET, given that we construct $\beta$ using simulations from the null model in both cases, we still obtain a fair comparison between all methods in terms of detection power, regardless of the choice of $\alpha$ for SUBSET and Bin-Weight.

%The justification for this arises from a consideration of the theoretical false alarm error rate under an aggregation of CUSUMs. (See, for example, Lemma 4 in the Supplementary Materials of \cite{SamworthWang18} for a result on the maximum of a univariate CUSUM procedure in the null setting.)

\subsection{Gaussian Setting, At Most One Change in Mean}\label{ssec:amocsims}

%Our first examination concerns the false alarm error rate of each of the methods. As stated, we fix this at 5\% for Bin-Weight, Inspect, Max and Mean. The penalty choices for SUBSET lead to no false positives across all simulated data scenarios ($n = \{1000, 10000, 100000\}$ and $d = \{5, 10, 50, 100, 500, 1000\}$).

To check the power of the methods in the single change setting we increase $\Delta$, the absolute change in mean - for each variate in which a change occurs - from 0.01 to 1.00 in increments of 0.01, and record the proportion of tests which yield a missed change in each case. We do this for $n = d = 1000$ and for densities of change corresponding to $0.5\%$, $1\%$, $5\%$, $10\%$, $50\%$ and $100\%$ of the variates affected by the change. 

\begin{figure}[!ht]
\begin{minipage}[t]{\textwidth}
\vspace{-10pt}
\begin{center}
\includegraphics[width=0.90\linewidth,keepaspectratio=true]{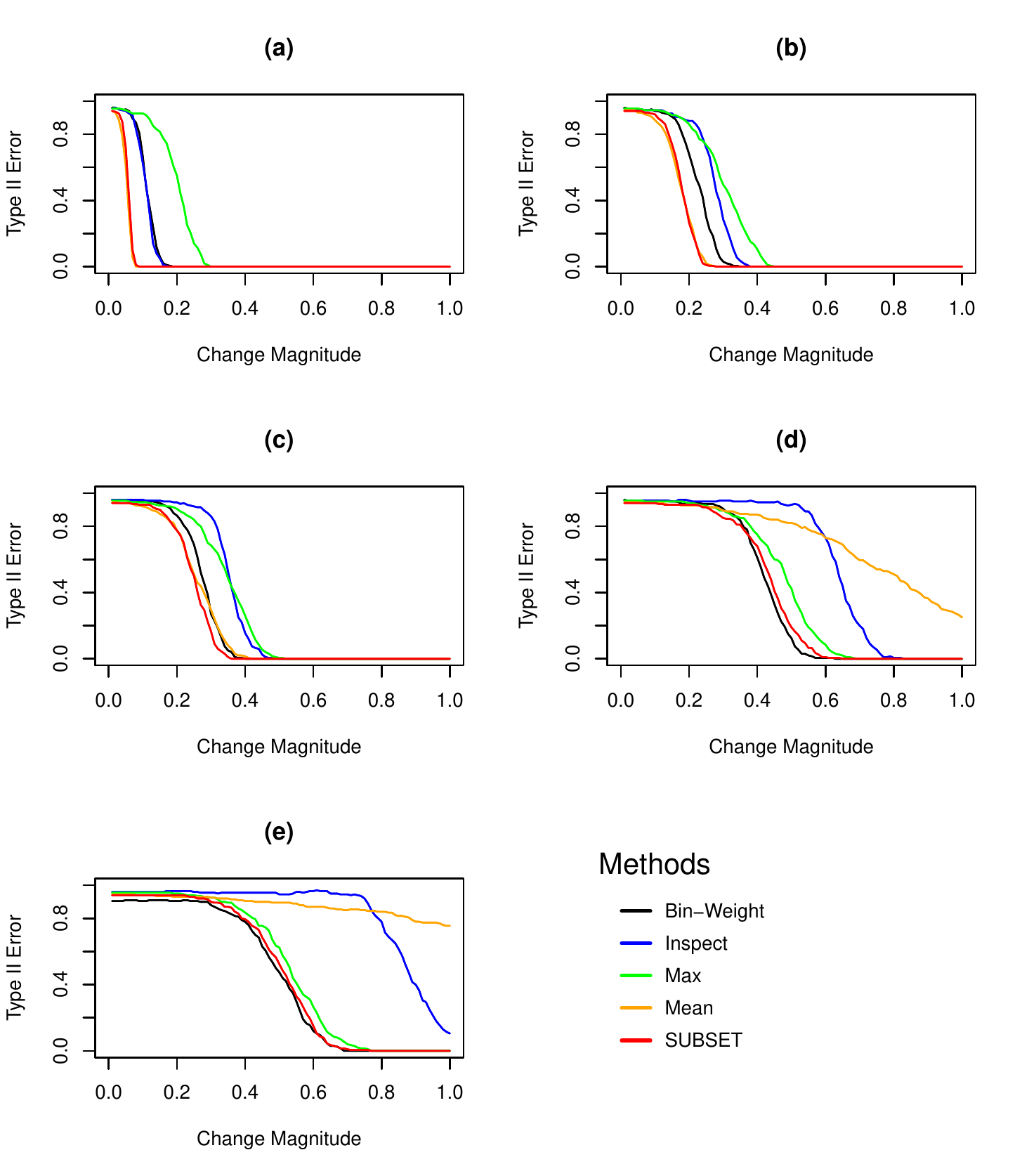}
\end{center}
\vspace{-40pt}
\begin{localsize}{10}
\caption{Type II Error probability (in the single change setting) across a range of values for $\Delta$ between 0.01 and 1 for each of the five methods under investigation for different subset densities of the changepoint, namely (a) 100\%; (b) 10\%; (c) 5\%; (d) 1\%; and (e) 0.5\%, keeping the temporal location of the changepoint fixed at $\tau = 50$ and $n = d = 1000$. Note that we aim to set the penalties for each method so that $\mathbb{P}($Type II Error$) \rightarrow 0.95$ when $\Delta \rightarrow 0$. }
\label{TypeII_1000variates}
\end{localsize}
\end{minipage}
\end{figure}

Figure~\ref{TypeII_1000variates} shows the result of this simulation when the location of the change is at $\tau = 50$, for each of the densities of change, and for each of the methods under investigation. Qualitatively similar results were observed for other settings; see, for example, Section B.4 of \cite{Tickle19}. The performance of different methods can be seen to depend on the proportion of series that change. For changes which affect a high proportion of variates, the Mean method is best, though SUBSET is very competitive. The other methods perform substantially worse, which is in keeping with the theory for these methods, which suggests that they are powerful for sparse changes but lose power for dense changes. By contrast, when the proportion of series that change is less than 1\%, the Mean method performs poorly. However SUBSET retains high power that is similar to or better than the competing methods. This is in accordance with Theorem~\ref{sufficiencycondition} which suggested that SUBSET would have good properties across both sparse and dense change scenarios.

We remark that it is surprising to see Inspect show much lower power than Max, SUBSET and Bin-Weight for the more sparse changes. We believe this is due to the default choice of tuning parameter used when performing the sparse projection of the data within the Inspect procedure. We found that increasing this tuning parameter, which leads to sparser projections, improves the performance of Inspect in these cases; note, however, that doing this will reduce the power for the cases where most series change. %For example, if the \verb|lambda| parameter is increased from its default of $\log\left(\frac{d\log(n)}{2}\right)$ to $\log(d \log n)$, we observe that, at a cost of increasing the value of $\Delta\mu$ required to give a negligible Type II Error in the case where the change affects all variates (around 0.3 as opposed to 0.2), the performance in sparse settings becomes much more competitive. (For example, in the 1\% density setting, the value of $\Delta\mu$ required to a give a negligible Type II Error is approximately $0.65$ as opposed to $0.80$.)

%{\color{red}{I am not sure if this is the most up to date plot, as we still observe slightly different false error rates for the methods for each plot. Also why is INSPECT so bad?}}

%This is the most up-to-date plot. Once the negative binomial experiments conclude, I'll re-run the INSPECT simulations without the bug which appears in the code for the package.

%We next compared the methods' computational speeds, examining the case where the affected set consisted of 50\% of variates (where possible - with 5 variates 3 of the variates underwent a change), and the change itself occurred at location $\theta = 0.184$ with $\Delta\mu = 1$. The resulting computation times for increasing $n$ and $d$, are shown in Table~\ref{timetaken} in Section~\ref{sec:simstudyadditional} in the Supplementary Materials. Table~\ref{timetaken} indicates that the SUBSET method does indeed scale approximately linearly in both $n$ and $d$, as do most of the competitor methods with the exception of Inspect. Note that the timings for the methods given in Table~\ref{timetaken} do not include the time taken to calculate the penalty values.

\begin{figure}[!ht]
\begin{minipage}[t]{\textwidth}
\vspace{-10pt}
\begin{center}
\includegraphics[width=0.90\linewidth, keepaspectratio=TRUE]{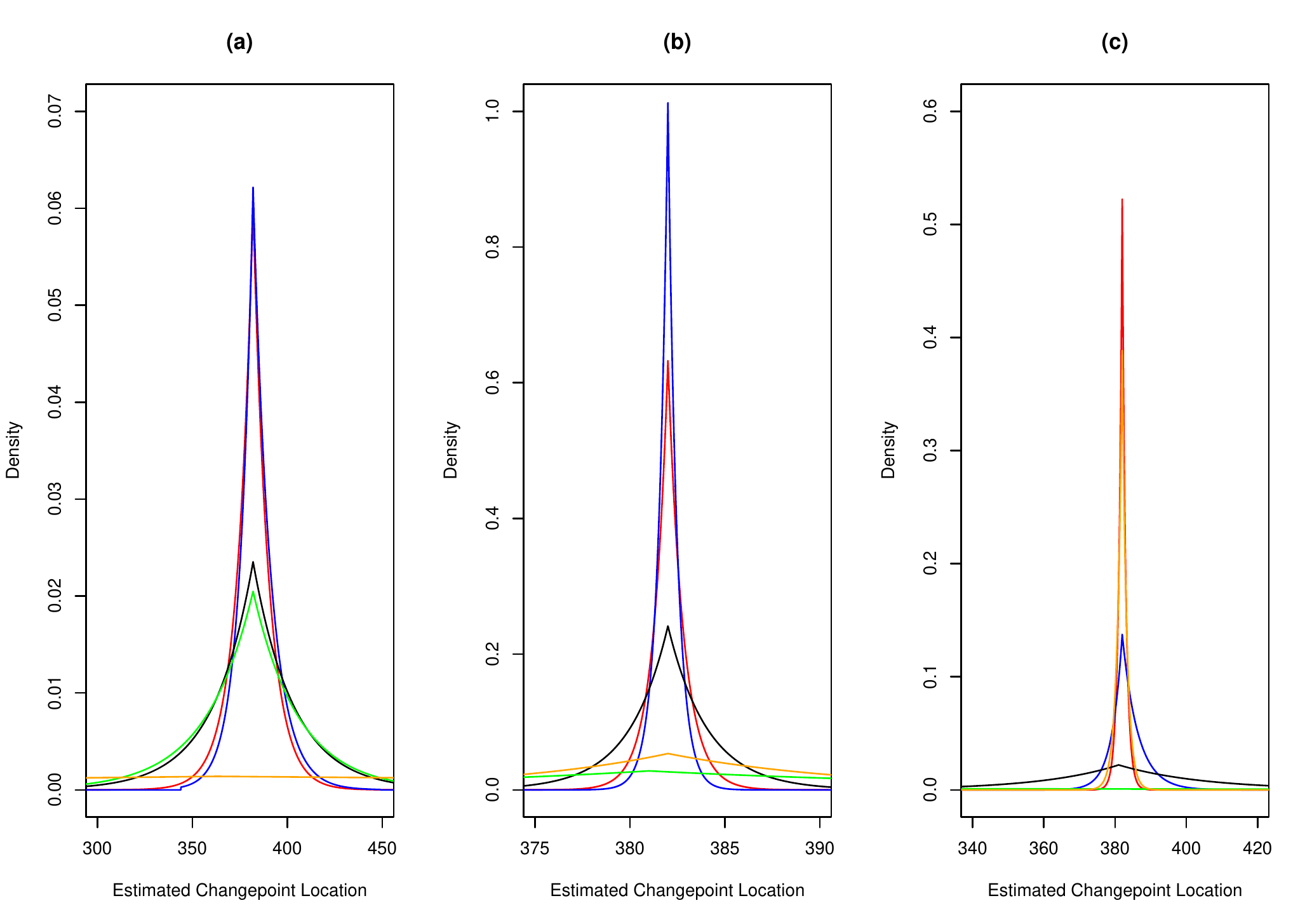}
\end{center}
\vspace{-40pt}
\begin{localsize}{10}
\caption{Estimated locations of the change (in the single change setting) under each of the five methods (Bin-Weight (black), Inspect (blue), Max (green), Mean (Orange) and SUBSET (red)) for (a) $|\mathcal{S}|/d = 0.005$ and $\Delta = 0.33$; (b) $|\mathcal{S}|/d = 0.05$ and $\Delta = 0.33$; and (c) $|\mathcal{S}|/d = 0.5$ and $\Delta = 0.1$, keeping $\tau = 382$ and $n = d = 1000$ in each case. The plots give the estimated densities, obtained using the log-concave density estimator of \cite{cule2009logconcdead}, for the change locations based on 200 repetitions under standard Gaussian noise.}
\label{LocationError}
\end{localsize}
\end{minipage}
\end{figure}

We next compare the average location errors of the methods for three scenarios, corresponding to a very sparse setting (0.5\% density of the change, and a change magnitude of $\Delta = 0.33$), a sparse setting (5\% density and $\Delta = 0.33$) and a dense setting (50\% density and $\Delta = 0.1$). We again consider $n = d = 1000$, as in Figure~\ref{TypeII_1000variates}, however we now have $\tau = 382$. The results are shown in Figure~\ref{LocationError}. For the two sparsest scenarios, we see that Inspect and SUBSET perform the best, with peaks correctly centred at the true changepoint. For the densest scenario, SUBSET demonstrates the best performance, with Mean a close competitor.

\begin{figure}[!ht]
\begin{minipage}[t]{\textwidth}
\vspace{-10pt}
\begin{center}
\includegraphics[width=0.90\linewidth, keepaspectratio=TRUE]{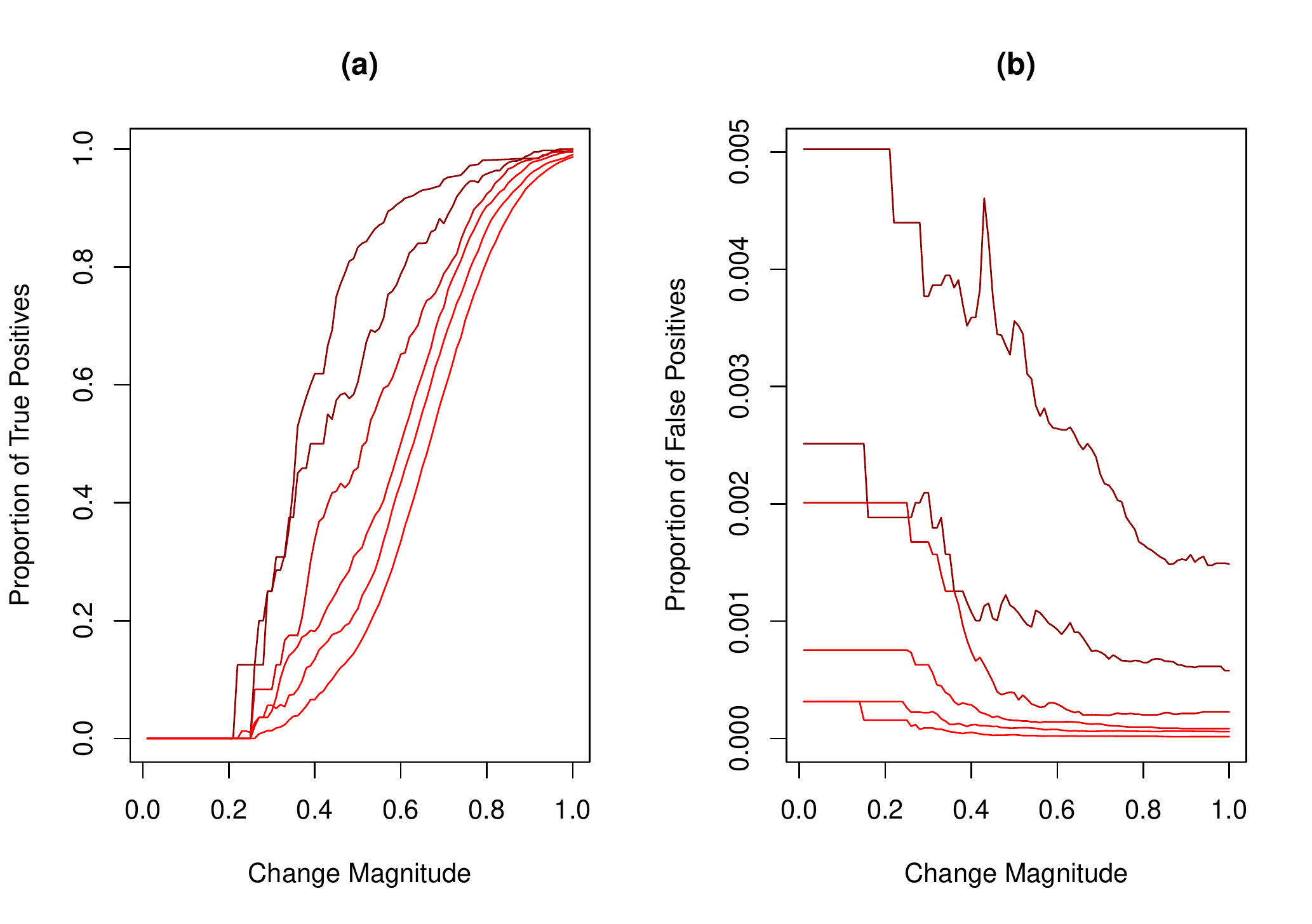}
\end{center}
\vspace{-40pt}
\begin{localsize}{10}
\caption{(a) The proportion of variates in affected set which are correctly identified as having been altered at the changepoint by SUBSET. (b) The proportion of variates not in the affected set which are correctly identified by SUBSET as having not been affected by the changepoint. The six scenarios (denoted from darkest red to lightest red) are $d = 200, 400, 800, 1600, 3200$ and $6400$, with $n = 400$ and $|\mathcal{S}|/d = 0.005$ in each case. Both plots are averages from 200 repetitions under standard Gaussian noise.}
\label{VariatePlots}
\end{localsize}
\end{minipage}
\end{figure}

The ability of SUBSET to estimate which subset of variates undergoes a change is also of interest. We examine the performance of SUBSET in determining the affected set in a setting with a small number of time points ($n = 400$) and increasing number of variates ($d = 200, 400, 800, 1600, 3200$ and $6400$). Given that the estimated affected set at a change is only returned by SUBSET if the procedure determines that putative changepoint is sparse, we take a very sparse setting, $|\mathcal{S}|/d = 0.005$. The results of this are shown in Figure~\ref{VariatePlots}, which indicates that SUBSET correctly identifies the affected set in sparse settings for sufficiently large $\Delta$. %In particular, we see that, for variates in the affected set at a sparse changepoint, SUBSET performs slightly better when $|\mathcal{S}|$ is smaller for smaller values of $\Delta$. This indicates that the $\alpha$ penalty potentially over-penalises changes which are sparse but which nevertheless may contain affected sets with more than a few variates. However, this difference becomes negligible for sufficiently large $\Delta$. 
Figure~\ref{VariatePlots} also indicates that SUBSET very rarely includes unaffected variates in the estimated affected set, with far fewer than 1\% of unaffected variates being incorrectly labelled in the worst cases, typically for small $\Delta$. %We observe that SUBSET performs worse under this metric if $|\mathcal{S}|$ is smaller, which highlights the tradeoff in the selection of an appropriate $\alpha$.

%We remark that in the setting where $(n, d, |\mathcal{S}|) = (400, 6400, 32)$, the location 

%Before examining settings with multiple changepoints, we remark on the properties of SUBSET in the very sparse setting. We retain $(n, d) = (400, 6400)$ from the previous example and take $|\mathcal{S}| = 2$ and $\Delta = $ INSERT VALUE.

\subsection{Gaussian Setting, Multiple Changes in Mean}\label{ssec:multisims}

We now compare methods for detecting multiple changepoints. We examine five scenarios, which we label as A, B, C, D and E here, each with three changepoints present. In each case, the changepoints may be found at times 600, 783 and 926. The only difference between scenarios is the size of each affected set of variates at each change. Thus, the scenarios imply different affected sets depending on the value of $d$. The scenarios are summarised below for $d = 1000$. Note that we once again fix $\sigma^2 = 1$ in all cases.
\begin{itemize}
\color{white}\item \color{black}A: All three changes affect all variates.
\color{white}\item \color{black}B: The first and third changes affect all variates; the second change affects 0.5\% of variates.
\color{white}\item \color{black}C: The first and third changes affect 0.5\% of variates; the second change affects all variates. 
\color{white}\item \color{black}D: All changes affect 1\% of variates.
\color{white}\item \color{black}E: The first, second and third changes affect 0.5\%, 1\% and 5\% of variates respectively.
\end{itemize}
Here, we restrict ourselves to examining the power of the methods. Note that we herein define a `missed change' as being a true changepoint for which the methods do not place an estimated change within $\lceil\log n\rceil$ points, while a `false alarm' is classed as an estimated changepoint for which no true changepoint is within $\left\lceil\log n\right\rceil$ time points. Note that the choice of a log-tolerance is motivated by classical theory on the localisation of changepoints in the univariate setting \cite[]{WangYuRinaldo20}. Each method was run within wild binary segmentation based on running each test on 1000 random intervals (except for Inspect, for which only 100 intervals were simulated due to the higher computational cost). The threshold used to detect changes was chosen such that for data with no change, each method had a false positive rate of 5\%. %For most of the methods, this simply involved collecting the value of the test statistic after running each procedure through a simulation from the null, which in turn consisted of drawing 1000 sub-intervals of the dataset uniformly at random. On each interval, the value of the test statistic was computed, and the maximum across all intervals returned. This was repeated 1000 times (for a different simulated null and a different set of random intervals), and the chosen penalty was taken as the 95\% empirical quantile of the resulting sample of returned values. For SUBSET, given the `two regime' nature of the test statistic, we compute the value of the test statistic under the sparse regime, and the value of the test statistic under the dense regime. Taking $K = \beta + d + \sqrt{2 \beta d}$, as per Theorem~\ref{allparameterchoices}, a value of $\beta$ for which either the sparse or dense statistic gives a change can be stored. We then take the 95\% empirical quantile of the resulting sample, as before. %{\color{red}{Sam: fill in details}}.

Table~\ref{multitypeii} shows the average number of changes - across 200 repetitions (100 for Inspect, again for computational reasons) - missed by each of the methods in each of the five scenarios for $n = d = 1000$ when $\Delta = 1$ for all variates which undergo a change at any changepoint. As can be seen from Table~\ref{multitypeii}, the best performing method in many scenarios was SUBSET, giving a low number of missed changes and false alarms in each case. In other words SUBSET performs well in both sparse and dense settings, even in cases with multiple changepoints. Only Inspect is competitive, with a slightly lower average number of missed changes in scenarios B and D, at the expense of a higher false alarm rate.
\begin{table}
\resizebox{\columnwidth}{!}{\begin{tabular}{ |p{3.3cm}||p{1.6cm}|p{1.6cm}|p{1.6cm}|p{1.6cm}|p{1.6cm}|}
 \hline
 \multicolumn{1}{|c||}{Average Number Missed} & \multicolumn{5}{|c|}{Method} \\
 \multicolumn{1}{|c||}{(Average False Alarms)} &\multicolumn{5}{|c|}{} \\
 \hline
  Scenario & SUBSET & Mean & Max & BW & Inspect \\
 \hline
A & \textbf{0.00} (\textbf{0.01}) & \textbf{0.00} (0.03) & 0.09 (1.06) & \textbf{0.00} (10.8) & \textbf{0.00} (0.32)\\ [-3pt]
\hline
B & 0.01 (\textbf{0.01}) & 0.99 (0.31) & 0.25 (1.05) & 0.99 (12.5) & \textbf{0.00} (0.32)\\ [-3pt]
\hline
C & \textbf{0.01} (\textbf{0.03}) & 1.98 (0.69) & 0.40 (1.21) & 1.99 (16.7) & 0.41 (0.32)\\ [-3pt]
\hline
D & 0.10 (\textbf{0.19}) & 2.50 (2.62) & 0.48 (1.21) & 2.97 (24.3) & \textbf{0.05} (0.42)\\ [-3pt]
\hline
E & \textbf{0.01} (\textbf{0.05}) & 1.76 (2.17) & 0.43 (1.09) & 2.79 (24.3) & 0.06 (0.68)\\
\hline
\end{tabular}
}
\begin{localsize}{10}
\caption{The average number of changes missed (and the average number of false alarms incurred) by each of the methods with $n = d =1000$ fixed in all cases and $\Delta = 1$ for any variate undergoing a change. Each of the scenarios A, B, C, D and E has 3 changepoints, and the percentage of variates affected by each change in each scenario is discussed at the beginning of Section~\ref{ssec:multisims}. Bold entries show the best performing algorithm.}
\label{multitypeii}
\end{localsize}
\end{table}

In order to suitably assess the performance of SUBSET in the relatively low-dimensional setting of the Global Terrorism Database, we conclude this section by examining a smaller
example.  %featuring the same changepoint locations as in scenarios A - E. 
Given that our treatment of the database involves converting the data to a 12-variate system, the following simulations involve $d = 12$. We define scenarios A' - D' as analogues to scenarios A - D, in which 
\begin{itemize}
\color{white}\item \color{black}A': All three changes affect all variates.
\color{white}\item \color{black}B': The first and third changes affect all variates; the second change affects the first and seventh variates.
\color{white}\item \color{black}C': The first and third changes affect only the first and seventh variates; the second change affects all variates. 
\color{white}\item \color{black}D': All changes affect only the first and seventh variates.
\end{itemize}

In addition, we include a ``surge" variant of each of these scenarios by adding two additional changepoints, at $\tau = 280$ and $\tau = 320$, which affect only the third variate. These two changes collectively form an epidemic change, so that the third variate returns to its original signal value.

Table~\ref{gaussiansmaller} shows the average number of missed changes and the average number of false alarms under 200 repetitions for each of the methods discussed, for $\Delta_i = 1$ at each changepoint - with $\Delta_3 = 5$ at the surge changepoints, if present - keeping $\sigma^2 = 1$ throughout. Once again, $n = 1000$, and the definitions of missed change and false alarm are as before, as are the constructions of the penalty values for each procedure.
\begin{table}
\resizebox{\columnwidth}{!}{\begin{tabular}{ |p{3.3cm}||p{1.6cm}|p{1.6cm}|p{1.6cm}|p{1.6cm}|p{1.6cm}|}
 \hline
 \multicolumn{1}{|c||}{Average Number Missed} & \multicolumn{5}{|c|}{Method} \\
 \multicolumn{1}{|c||}{(Average False Alarms)} &\multicolumn{5}{|c|}{} \\
 \hline
  Scenario & SUBSET & Mean & Max & BW & Inspect \\
 \hline
A', Surge & 0.01 (0.10) & \textbf{0.00} (0.27) & 0.18 (0.32) & 0.01 (\textbf{0.07}) & 0.01 (0.10)\\
\hline
A', No Surge & 0.01 (0.02) & \textbf{0.00} (\textbf{0.01}) & 0.19 (0.26) & 0.01 (0.04) & 0.01 (0.10) \\
 \hline
B', Surge & \textbf{0.06} (0.15) & 0.38 (0.67) & 0.24 (0.36) & \textbf{0.06} (\textbf{0.13}) & \textbf{0.06} (\textbf{0.13})\\
\hline
B', No Surge & \textbf{0.06} (\textbf{0.08}) & 0.38 (0.41) & 0.24 (0.30) & 0.07 (0.10) & \textbf{0.06} (0.13) \\
\hline
C', Surge & 0.17 (0.21) & 0.89 (0.99) & 0.30 (0.33) & 0.27 (0.27) & \textbf{0.14} (\textbf{0.18}) \\
\hline
C', No Surge & \textbf{0.09} (\textbf{0.10}) & 0.73 (0.57) & 0.31 (0.35) & 0.24 (0.20) & 0.13 (0.18)\\
 \hline
D', Surge & 0.22 (0.26) & 1.27 (1.38) & 0.37 (0.40) & 0.32 (0.32) & \textbf{0.20} (\textbf{0.23})\\
\hline
D', No Surge & \textbf{0.14} (\textbf{0.16}) & 1.10 (0.95) & 0.38 (0.41) & 0.28 (0.25) & 0.19 (0.22) \\
 \hline
\end{tabular}
}
\begin{localsize}{10}
\caption{The average number of changes missed (and the average number of false alarms incurred) by each of the methods with $n = 1000$ fixed in all cases. Each of the scenarios A' - D' are as detailed above. If a surge a present, there are two additional sparse changes positioned close together which collectively have no impact on the mean. Bold entries show the best performing algorithm with respect to each measure.}
\label{gaussiansmaller}
\end{localsize}
\end{table}

We observe from Table~\ref{gaussiansmaller} that all methods (with the possible exception of Max) do fractionally worse in the presence of the ``surge" segment. However performance remains roughly consistent. In the case of SUBSET, we see that both the average number of missed changes and the average number of false alarms incurred remain very low, with only Inspect being a consistent competitor across all scenarios. 

%We observe from Table~\ref{gaussiansmaller} that the performance of all methods is markedly worse under these smaller examples than for the more data-rich examples given in Table~\ref{multitypeii}. This is particularly true in the $\Delta = 1$ case, in which all methods miss around half of the true changepoints present, but in which SUBSET, Max and Inspect appear to have marginally better statistical power due to the "sparsity" of most of the changepoints present. For the $\Delta = 5$ setting, all methods perform much better, in particular Inspect.

\subsection{Negative Binomial Setting}\label{ssec:negbinsims}

Finally, motivated by problem of detecting changes in the global terrorism data, we consider detecting changes in count data. To allow for over-dispersion relative to a Poisson model, we assume that each data point is the realisation of a negative binomial random variables.
Formally, we have
\begin{equation}\label{negbinmodel}
%y_{i,j} \sim \begin{cases} \mbox{Neg-Bin}\left(r_{i,1}, p_{i,1} \right) \mbox{ for } 1 \leq j \leq \tau_1, \\ \mbox{Neg-Bin}\left(r_{i,2}, p_{i,2} \right) \mbox{ for } \tau_2 + 1 \leq j \leq \tau_2, \\ \ldots \\ \mbox{Neg-Bin}\left(r_{i,m+1}, p_{i,m+1} \right) \mbox{ for } \tau_m + 1 \leq j \leq n\end{cases} \mbox{ for } i \in \left\{1, \ldots, d\right\},
y_{i,j} \sim \mbox{Neg-Bin}\left(r_{i}, p_{i,k} \right), \mbox{ for } \tau_{k-1}+1 \leq j \leq \tau_k
\end{equation}
for a common unknown over-dispersion parameter, $r_i$
%for some sequence, $\left(r_{i,k} \right)_{k=1}^{m+1}$, of unknown over-dispersion parameters 
and some sequence, $\left(p_{i,k}\right)_{k=1}^{m+1}$, of unknown success probabilities. 

At the time of writing, we believe that our method is the only one which naturally extends to such a setting, as we can re-define the likelihood ratio test statistic that our procedure is based on so that it relates to the negative binomial model we are fitting.  In particular, we use the likelihood ratio test statistic for a change in success probability, assuming a common over-dispersion parameter, with the over-dispersion parameter estimated by a methods of moments estimator \citep{SavaniZhigljavsky06}. This test is a natural one for detecting changes in the mean of the data. Note that we use a method of moments estimator, rather than the maximum likelihood estimator, given the difficulty of computing the latter to appropriate precision in efficient time under a negative binomial model.

We evaluate the performance of SUBSET for scenarios similar to the low-dimensional examples discussed in Section~\ref{ssec:multisims}. 
In the examples here, we take a change in probability of 0.1 at each variate which is affected at each changepoint. We take three different over-dispersion parameters of $r = 3, 20$ and $100$, to give a total of twenty-four different experiments, with half of the experiments including a surge at the same location as in the Gaussian example, in which there is an epidemic change of size $p = 0.1$ for the third variate only. The results are shown in Table~\ref{negbinsmaller}.
\begin{table}
\resizebox{\columnwidth}{!}{\begin{tabular}{ |p{3.3cm}||p{1.6cm}|p{1.6cm}|p{1.6cm}|p{1.6cm}|p{1.6cm}|p{1.6cm}|p{1.6cm}|p{1.6cm}|}
 \hline
  Scenario & \multicolumn{2}{|c|}{A'} & \multicolumn{2}{|c|}{B'} & \multicolumn{2}{|c|}{C'} & \multicolumn{2}{|c|}{D'} \\
  \multicolumn{1}{|c||}{} & Surge & No Surge & Surge & No Surge & Surge & No Surge & Surge & No Surge \\
 \hline
 \multicolumn{1}{|c||}{Over-Dispersion} & \multicolumn{8}{|c|}{Average Number Missed} \\
 \multicolumn{1}{|c||}{} & \multicolumn{8}{|c|}{(Average False Alarms)} \\
 \hline
$r=3$ & 0.07 (0.09) & 0.06 (0.08) & 0.48 (0.24) & 0.52 (0.24) & 1.04 (0.35) & 0.96 (0.36) & 1.58 (0.79) & 1.59 (0.79)\\
\hline
$r=20$ & 0.00 (0.00) & 0.00 (0.00) & 0.01 (0.01) & 0.01 (0.01) & 0.01 (0.01) & 0.01 (0.01) & 0.04 (0.04) & 0.04 (0.05)\\
\hline
$r=100$ & 0.00 (0.00) & 0.00 (0.00) & 0.00 (0.00) & 0.00 (0.00) & 0.00 (0.00) & 0.00 (0.00) & 0.00 (0.00) & 0.00 (0.00) \\
\hline 
\end{tabular}
}
\begin{localsize}{10}
\caption{The average number of changes missed, and the average number of false alarms incurred, by SUBSET with $n = 1000$ fixed in all cases, and $\Delta p = 0.1$. Each of the scenarios is detailed in Section~\ref{ssec:multisims}, and a surge corresponds to a short period with a low $p$ in the third variate only.}
\label{negbinsmaller}
\end{localsize}
\end{table}

We see from Table~\ref{negbinsmaller} that SUBSET consistently performs well in the small $d$ settings, with performance improving for larger values of the over-dispersion parameter $r$ and fewer sparse changes present. Note that in the low $r$ setting, SUBSET misses a significant proportion of sparse changes if most of the changes are sparse. This suggests that, for those periods of time and geographical regions where the number of terrorist incidents is low, there is a non-trivial probability that SUBSET will fail to detect changes in the latent probability of a terrorist attack.

\section{Detecting Changes in Global Terrorism}\label{sec:real}

We now return to the Global Terrorism Database, and apply SUBSET to detect changes in the recorded incidence of terrorism events. Of particular interest is whether our method can detect geographically-localised changes, such as those that have previously been noted in the discussion in Section~\ref{sec:gtdintro}, as well as changes that affect all series, for example due to changes in how events are recorded in the database.

One approach to modelling the terrorism count data might be to adopt a Poisson likelihood, such that a changepoint corresponds to an alteration in the rate parameter. However, we found that the data poorly conformed to a Poisson,  due to a high degree of over-dispersion, and the use of a Poisson-based segment cost function led to SUBSET placing a very large number of changepoints. We therefore model the data for each series as realisations from a negative binomial with changing `success' probability parameter, and apply SUBSET as per the study in Section~\ref{ssec:negbinsims}.  %The results of our analysis are displayed in Table~1 and Figure~1 of the Supplementary Material. These document the months in which the estimated changepoints of the period occurred and the corresponding geographical locations affected. 

%Indeed, it is now natural to assume that, for a given univariate series within the stream corresponding to a particular region, say $\left(Y_{i, j}\right)_{j=1}^{564}$:

%\begin{equation}\label{negbinmodel}
%Y_{i,j} \sim \begin{cases} \mbox{Neg-Bin}\left(r_1, p_1 \right) \mbox{ for } 1 \leq j \leq \tau_1, \\ \mbox{Neg-Bin}\left(r_2, p_2 \right) \mbox{ for } \tau_2 + 1 \leq j \leq \tau_2, \\ \ldots \\ \mbox{Neg-Bin}\left(r_{m+1}, p_{m+1} \right) \mbox{ for } \tau_m + 1 \leq j \leq 564 \end{cases} \mbox{ for } i \in \left\{1, \ldots, 12\right\},
%\end{equation}

%for some sequence, $\left(r_{k} \right)_{k=1}^{m+1}$, of unknown over-dispersion parameters and some sequence, $\left(p_{k}\right)_{k=1}^{m+1}$, of unknown success probabilities. Given the difficulty of computing the maximum likelihood estimators for the former at a given stage of the procedure, we assume that the over-dispersion parameter changes only if the unknown success probability also changes, and we compute a methods of moments estimator for these over-dispersion parameters at each stage of the procedure for the purposes of computation time.

\begin{figure}[!ht]
\begin{minipage}[t]{\textwidth}
\begin{center}
\includegraphics[width=0.80\linewidth,keepaspectratio=true]{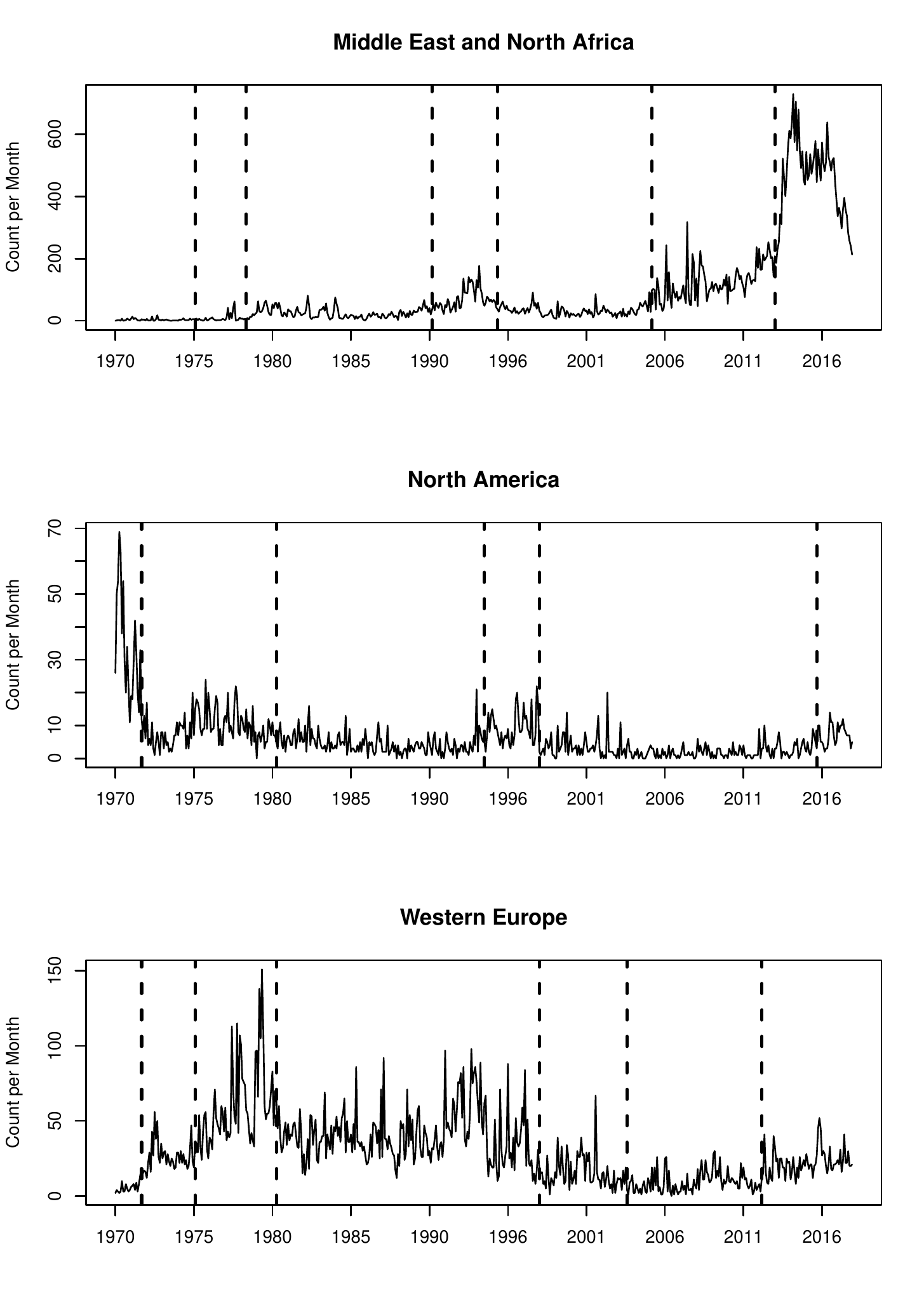}
\end{center}
\vspace{-20pt}
\begin{localsize}{10}
\caption{Terrorism incident count per month for the Middle East and North Africa (top), North America (middle) and Western Europe (bottom) from January 1970 to December 2017. Changes found by the SUBSET method using a negative binomial cost function are overlaid as dashed vertical lines.}
\label{threeplots}
\end{localsize}
\end{minipage}
\end{figure}

Figure~\ref{threeplots} displays the output of our analysis for the Middle East and North Africa, North America and Western Europe regions, highlighting the dates of the changes which affect these regions. For a fuller picture, with all twelve regions displayed, please see Section~\ref{ssec:subsetgtd} of the Supplementary Material.

Some notable features are apparent. For example, one of the very few dense changes located by SUBSET (in January 1998) corresponds to an alteration in the data collection method for all regions. As noted in Section~\ref{sec:gtdintro}, between 1970 and 1997, the Pinkerton Global Intelligence Services collated the information for the database from international reports in effectively `real-time'. From the beginning of 1998, events were only added to the database retrospectively by the Study of Terrorism and Responses to Terrorism and the Center for Terrorism and Intelligence Studies. In addition, the definition of a terrorist event was refined into a set of six criteria, with an event having to satisfy at least five of the criteria to be included in the database. %For more information, please see the START website which hosts the GTD as well as \cite{LaFreeDugan07}. 
The overall effect, however, was an expansion in the definition of what constitutes a terrorist action. In this context, a dense changepoint is unsurprising. We remark that the Middle East and North Africa region is currently one of the few regions which is not labelled as affected by the dense change at the beginning of 1998. This is due to the post-processing step within SUBSET. 

The other changepoints which affect more than one or two of the series in our analysis are concentrated at the end of the 1970s. This again conforms to the general picture discussed by \cite{LaFreeDugan07} and \cite{LaFree10} and highlighted in Section~\ref{sec:gtdintro}. Eight of the twelve regions are affected by changepoints between 1975 and 1978, with Central America and the Caribbean and Western Europe particularly hard-hit by the increase in terrorism during this time period. In the case of Western Europe, as discussed in Section~\ref{sec:gtdintro}, the change seen in February 1975 appears to be due to a change of British government policy in Northern Ireland. One possible policy (which aligns exactly with the changepoint) is the truce between the Provisional IRA and the British Army, which led to an increase in sectarian violence and subsequent retaliations; see, for example, \cite{Craig14} and \cite{White10}. We remark that the dense changepoint in 1998 also aligns closely to the Good Friday Agreement. Meanwhile, the changes in Central America during the same time period seemingly can be explained by the appearance of revolutionary groups in the late 1970s. The activities of these groups waned drastically as countries in the region democratised into the 1990s, a process often referred to in the Central American context as Democratic Consolidation; see, for example, \cite{Berntzen93}.

Otherwise, as expected, most of the estimated changes are sparse. This corresponds to the commentary found in, for example, \cite{LaFreeDuganMiller14}, which asserts that most causes of terrorism remain localised. For instance, the change in the Middle East and North Africa in early 2013 appears to correspond to the beginning of the so-called `Arab Winter' \cite[]{King20, Mihaylov17}, a term which is used to describe the still ongoing re-emergence of significant levels of authoritarianism and extremism following the Arab Spring. Other changes of interest in the series include those placed in September 1978 and March 2005. We note that the former date aligns with the signing of the Camp David Accords. While the Accords did much to stabilise relations between Israel and Egypt, tensions were also engendered between Egypt and other Arab nations, as well as between the Egyptian government and their people \cite[]{BaniSalamahetal12, Quandt86, Quandt88}. Indeed, President Anwar Sadat was himself assassinated in 1981. The latter date detected by SUBSET is around one year before the conclusion of the insurgency in Iraq which degenerated into sectarian violence in February 2006. 

% Meanwhile the period of more intense activity in Western Europe in the later 1970s seems to broadly align with some of the worst years of the Troubles. The full results for all twelve regions can be found in Section~4 of the Supplementary Material (Table~1 and Figure~1) These document the months in which the estimated changepoints of the period occurred and the corresponding geographical locations affected. 

%\section{Conclusions}\label{sec:discussion}

%We have proposed SUBSET, a means of computationally efficient multivariate changepoint detection. This method incorporates a penalised likelihood appraoch with that of Wild Binary Segmentation. We have demonstrated that the method has good theoretical and computational properties in a variety of cases, including the at most one change in mean problem to multiple change problems which potentially exhibit more complex behaviour at each changepoint, such as those seen in the Global Terrorism Database. 

Our analysis of the Global Terrorism Database fundamentally shows that the assertion that terrorist activity remains localised is broadly correct. By distinguishing between sparse and dense changepoints, SUBSET is able to confirm that most ``dense'' changepoints in the context of global terror levels in fact match up to changes in data collection procedures, as well as the fluidity in the definition of terrorism itself. By contrast, the sparse changes found by SUBSET appear to align with political events of particular significance in regions of greater tension. %Therefore, given that terrorism worldwide is now much more prevalent than at the genesis of the Global Terrorism Database, historical insights gained from a regional and national level are increasingly prescient. 

%We have proposed SUBSET, a means of computationally efficient multivariate changepoint detection. This method incorporates a penalised likelihood approach with that of Wild Binary Segmentation. We have demonstrated that the method has good theoretical and computational properties in a variety of cases. These cases range from the at most one change in mean problem to more complex multiple change problems which potentially exhibit more difficult behaviour at each changepoint. In addition, we believe that the suggestions for implementation made here, such as the appropriate settings of the penalty values, will be of use to practitioners. 

%Some remaining challenges include an explicit algorithmic treatment of correlated or lagged changes to provide a clearer quantitative picture of a common cause of a change. Presently, this is an issue of penalty adjustment. Another issue to overcome is the fact that this method is best employed under specific parametric assumptions. It would be desirable to find a setting for this method under which these may be relaxed. Perhaps the most important issue from a data streaming perspective, however, is that this method, while efficient, is highly offline. 

%It is this latter challenge in particular, namely achieving a reliable sequential changepoint detection algorithm in a general high-dimensional setting, that we believe forms the basis of the most interesting problem arising from this method.

\section{Conclusions}\label{sec:discussion}

We have proposed SUBSET, a means of computationally efficient multivariate changepoint detection. This method incorporates a penalised likelihood approach with that of Wild Binary Segmentation. We have demonstrated that the method has good theoretical and computational properties in a variety of cases. These cases range from the at most one change in mean problem to more complex multiple change problems which potentially exhibit more difficult behaviour at each changepoint.  An important feature of SUBSET is that it is likelihood based, and thus can be applied to various data-types, and even situations where we wish to multiple data streams of mixed type (continuous, count, categorical etc.).

One of the main limitations of SUBSET, as with most current change detection methods, is that it assumes independence across data streams and time. The method can be made robust to violations of these assumptions through adjusting the penalty function for adding a changepoint, albeit with a potential loss of power. It would be desirable to extend the modelling approach that underpins SUBSET so as to relax these independence assumptions, perhaps building on ideas in \cite{Tvetenetal20} or \cite{ChoFryzlewicz21}.

\section{Acknowledgments}
 This paper is based on work completed while Tickle was part of the EPSRC funded STOR-i Centre for Doctoral Training (EP/L015692/1). Eckley and Fearnhead gratefully acknowledge the financial support of EPSRC grant EP/N031938/1. The authors also acknowledge British Telecommunications plc (BT) for financial support, and are grateful to Kjeld Jensen, Dave Yearling and Guillem Rigaill for helpful discussions. Finally, the authors would like to thank the editors and two anonymous reviewers for several helpful comments and feedback. 

\bibliographystyle{royal}
\bibliography{References.bib}

\pagebreak

\begin{center}
    \LARGE{A computationally efficient, high-dimensional multiple changepoint procedure with application to global terrorism incidence} \\
    \Huge{Supplementary Material} \\
\end{center}

\setcounter{section}{0}
\appendix

\section{Preliminary Lemmas}\label{sec:prelimlemmas}

In this section, we establish several lemmas required to prove the central results of the main article (please see Section~\ref{sec:proofs} for the main proofs). Our general approach with this section is to establish the stated results in either the sparse or the dense setting, and then combine these results appropriately in Section~\ref{sec:proofs}. Throughout, we repeatedly use the following two lemmas: 

\begin{lemma}[\citet{LaurentMassart00}]\label{MassartLaurent}
Suppose $G \sim \chi^2_k$. Then for any $x > 0$
\[
\mathbb{P}\left(G \geq k + 2 \sqrt{xk} + 2x\right) \leq \exp\left(-x\right).
\]
\end{lemma}

%\textbf{Proof}: See, for example, \cite{LaurentMassart00}.

\begin{lemma}[\cite{Birge01}]\label{BirgeMassart}
Suppose $H \sim \chi^2_k(\nu)$. Then for any $y > 0$
\[
\mathbb{P}\left(H \geq k + \nu - 2 \sqrt{(k + 2\nu)y}\right) \geq 1 - \exp(-y),
\]
and
\[
\mathbb{P}\left(H \geq k + \nu + 2 \sqrt{(k + 2\nu)y} + 2y\right) \leq \exp(-y).
\]
\end{lemma}

%\textbf{Proof}: See \cite{Birge01}.

We now give two results on the Type I error of the SUBSET procedure. Lemma~\ref{typeIsparse} gives a bound on the Type I error in the sparse setting, under particular choices for the penalties $\alpha$ and $\beta$, and Lemma~\ref{typeIdense} gives an equivalent result in the dense setting, under an additional choice for the dense penalty $K$.

\begin{lemma}\label{typeIsparse}
Suppose we are in the same setting as for Theorem~\ref{allparameterchoices} of the main article. Let $D_{i,t}^{'}$, $\alpha$ and $\beta$ be as defined in Section~\ref{sec:themethod} of the main article. Define $S_{1,t} = \sum_{i=1}^d D_{i,t}^{'} - \beta$. Let $\sqrt{\beta} = \sqrt{2d\frac{\Gamma\left(\frac{1}{2}, \frac{\alpha}{2}\right)}{\Gamma\left(\frac{1}{2}\right)}} + C \sqrt{\log n}$ for some constant $C$, then
\[
\mathbb{P}\left(\max_t S_{1,t} > 0\right) \leq n^{1 - \frac{C^2}{2}} \exp\left(\frac{(1 + \vartheta)^2}{4 \vartheta}\frac{d}{\exp\left(\frac{\alpha}{2}\right)}(1 + \alpha)^{-\frac{1}{2}}\right), \text{ some } \vartheta,
\]
providing that 
\begin{equation}\label{alphacondition}
\frac{\gamma\left(\frac{1}{2},\frac{\alpha}{2}\right)}{\Gamma\left(\frac{1}{2}\right)} \geq \vartheta > 0, 
\end{equation}
and
\begin{equation}\label{betacondition}
\beta > 2 d \frac{\Gamma\left(\frac{1}{2},\frac{\alpha}{2}\right)}{\Gamma\left(\frac{1}{2}\right)}.
\end{equation}
\end{lemma}

\begin{lemma}\label{typeIdense}
Suppose again that we are in the same setting as for Theorem~\ref{allparameterchoices} of the main article. Let $D_{i,t}$ and $K$ be defined as in Section~\ref{sec:themethod} of the main article. Define $S_{2,t} = \sum_{i=1}^d D_{i,t} - K$. Setting $\beta = \left(J + \epsilon\right) \log n$, $\alpha = 2 \log d$ and $K = d + \sqrt{2 \beta d} + \beta$ gives that
\[
\mathbb{P}\left(\max_t S_{2,t} > 0\right) \leq n^{1 - \frac{\left(J + \epsilon\right)}{2}}.
\]
\end{lemma}

\textbf{Proof of Lemma \ref{typeIsparse}}: Fix $\tau$. Note if $f_{D_{i,\tau}^{'}}(x)$ is the density function for $D_{i, \tau}^{'}$ then it is straightforward to show that for $x > 0$, $d \log f_{D_{i, \tau}^{'}}(x)/dx < -\frac{1}{2}$; therefore, $D_{i, \tau}^{'}$ is stochastically dominated by $N_{i,\tau}$, where 
\[
N_{i, \tau} = 
\begin{cases}
0 &\mbox{ w.p. } p_{\alpha}\\
\mbox{Exp}\left(\frac{1}{2}\right) &\mbox{ w.p. } 1 - p_{\alpha},
\end{cases}
\]
such that $p_{\alpha} = \mathbb{P}\left(D_{i,\tau}^{'} = 0\right) = \frac{\gamma\left(\frac{1}{2}, \frac{\alpha}{2}\right)}{\Gamma\left(\frac{1}{2}\right)}$. We define for convenience $q_{\alpha} = 1 - p_{\alpha} = \frac{\Gamma\left(\frac{1}{2}, \frac{\alpha}{2}\right)}{\Gamma\left(\frac{1}{2}\right)}$.

Let $A_{\tau} = \sum_{i=1}^d N_{i,\tau}$; then the moment generating function of $A_{\tau}$ is
\[
m_{A_{\tau}}\left(\lambda\right) = \left(p_{\alpha} + \frac{q_{\alpha}}{1 - 2 \lambda}\right)^{d}.
\]
We seek the Cramer transform, $\psi_{A_{\tau}}^{*}(r)$, of $A_{\tau}$, such that
\[
\psi_{A_{\tau}}^{*}(r) = \sup_{\lambda \geq 0} \left\{\lambda r - d \log \left(p_{\alpha} + \frac{q_{\alpha}}{1 - 2 \lambda} \right) \right\};
\]
for $\lambda < \frac{1}{2}$ the supremum is achieved close to $\lambda^{'} = \frac{1}{2p_{\alpha}} - \frac{1}{p_{\alpha}} \sqrt{\frac{d q_{\alpha}}{2r}}$, particularly for large $d$, and we can bound the Cramer transform by the value of the argument at $\lambda'$. Thus we have
\begin{align*}
\mathbb{P}\left(A_{\tau} \geq \beta\right) &\leq \exp\left(-\psi_{A_{\tau}}^{*}(\beta)\right) \\
&\leq \exp\left(-\left(\frac{1}{2 p_{\alpha}} - \frac{1}{p_{\alpha}} \sqrt{\frac{d q_{\alpha}}{2\beta}} \right) \beta \right) \left(\frac{p_{\alpha}}{1 - \sqrt{\beta q_{\alpha}}{2 d} }\right)^d \\
&\leq \exp\left(-\frac{1}{2p_{\alpha}} \left(\sqrt{\beta} - \left(1 + p_{\alpha}\right) \sqrt{\frac{d q_{\alpha}}{2}}\right)^2\right) \exp\left(\frac{(1 + p_{\alpha})^2 d q_{\alpha}}{4 p_{\alpha}} + d \log p_{\alpha} \right) \\
&\leq \exp\left(\frac{(1 + \vartheta)^2}{4 \vartheta} Q \right) \exp\left(-\frac{1}{2}\left(\sqrt{\beta} - \sqrt{2Q}\right)^2\right),
\end{align*}
for $Q = d q_{\alpha}$, where the penultimate line follows from considering $F$ such that $\left(1/p_{\alpha} - \frac{1}{p_{\alpha}} \sqrt{\frac{\beta q_{\alpha}}{2d}}\right)^{-d} \leq \exp(\sqrt{\beta} F)$ and performing a Taylor Series expansion, and the final line follows from conditions~(\ref{alphacondition}) and~(\ref{betacondition}). 

Let $\sqrt{\beta} = \sqrt{2Q} + C\sqrt{\log n}$, for some $C$. We now use the fact that $\frac{\Gamma\left(v, w\right)}{\Gamma\left(v\right)} \leq e^{-w} \left(1 + \frac{w}{v} \right)^{v - 1}$ (which can be shown using Jensen's Inequality), to assert that $ q_\alpha \leq e^{-\alpha/2} (1 + \alpha)^{-1/2}$, and that therefore
\[
\mathbb{P}\left(A_{\tau} \geq \beta \right) \leq n^{-\frac{C^2}{2}} \exp\left(\frac{(1 + \vartheta)^2}{4\vartheta}\frac{d}{\exp\left(\frac{\alpha}{2}\right)}\left(1 + \alpha\right)^{-\frac{1}{2}}\right);
\]
performing a Bonferroni correction for the position of $\tau$ in the data then gives the stated result. \QEDB

\textbf{Proof of Lemma~\ref{typeIdense}}: In the scenario where there is no true change, the difference in cost between selecting the point $\tau$ as a change (with affected subset $\mathcal{S} = \left\{1, \ldots, d\right\}$) and simply finding the (correct) null model is
\begin{align*}
\mbox{Diff} &= \mbox{RSS} \left(\textbf{y}_{1:n} ; \emptyset \right) - \mbox{RSS} \left(\textbf{y}_{1:n} ; \tau; \mathcal{S} \right) - K \\
&:= W - K,
\end{align*}
where here we use the notation $RSS(\textbf{z}; \xi; \mathcal{T})$ to denote the residual sum of squares of the vector $\textbf{z}$ enforcing a changepoint at time $\xi$ with affected set $\mathcal{T}$. Note that $W \sim \chi^2_d$. By Lemma~\ref{MassartLaurent}, to establish the result we require $K$ and $x$ such that
\begin{align*}
K &= d + 2 \sqrt{x d} + 2 x \\
\exp\left(-x\right) &= n^{-\frac{J}{2} - \epsilon/2},
\end{align*}
giving $x = \left(J + \epsilon\right)/2 \log n = \beta/2$, and consequently $K = d + \sqrt{2\beta d} + \beta$ as required. \QEDB

We later use these lemmas to establish Theorem~\ref{allparameterchoices}. Before this, we give further results which are needed in establishing the other result of the main article.

\begin{lemma}\label{sparsepower}
Assume that we are in the same setting as for Lemma~\ref{typeIsparse}, except now we have that $\mu_{i,1} \neq \mu_{i,2}$ whenever $i \in \mathcal{S} \subseteq \left\{1, \ldots, d \right\}$. For $i \in \mathcal{S}$, let $\Delta_i := \left|\mu_{i,2} - \mu_{i,1}\right|$. Then for $\delta > 0$ and $a = \max\left\{n, d \right\}$ a sparse changepoint will be detected by SUBSET with probability greater than $1 - \left(a \right)^{-\delta}$, providing that
\[
\sum_{i \in \mathcal{S}} \left(\Delta_i\right)^2 \geq \frac{4 \delta \log a + \beta + \left|\mathcal{S}\right|(\alpha - 1) + 2 \sqrt{\delta \log a \left((2\alpha - 1)|\mathcal{S}| + 2 \beta + 4 \delta \log a \right)}}{n \theta \left(1 - \theta\right)},
\]
where here we have $\theta = \frac{\tau}{n}$ is fixed strictly between $0$ and $1$.
\end{lemma}

\begin{lemma}\label{densepower}
Assume that we are in the same setting as for Lemma~\ref{sparsepower}, except with the dense penalty regime. Then, again with probability greater than $1 - a^{-\delta}$, for $ 2 > \delta > 0$ and $a = \max \{n, d\}$, providing that
\[
\sum_{i=1}^d \left(\Delta_i\right)^2 \geq \frac{4 \delta \log a + K - d + 2 \sqrt{\delta \log a \left(4 \delta \log a + 2 K - d\right)}}{n \theta \left(1 - \theta\right)},
\]
a changepoint will be detected in the dense setting.
\end{lemma}

\textbf{Proof of Lemma~\ref{sparsepower}}: Suppose there is a true change at location $\tau$ which affects a non-empty, sparse subset $\mathcal{S} \subset \left\{ 1, \ldots, d\right\}$ of variates, such that the magnitude of change in variate $i$ is $\Delta_i$. We compare the cost of fitting no change in such a scenario against the cost of fitting the truth; i.e. let
\[
\mbox{Diff} := \sum_{i \in \mathcal{S}} D_{i,\tau} - \beta - |\mathcal{S}|\alpha,
\]
where $D_{i,\tau}$ is as defined in the main article. Note that $D_{i,\tau} \sim \chi^2_1\left(n\theta\left(1-\theta\right)(\Delta_i)^2\right)$, so 
\[
\mbox{Diff} + \beta + |\mathcal{S}|\alpha \sim \chi^2_{|\mathcal{S}|}\left(n\theta\left(1 - \theta\right) \sum_{i \in \mathcal{S}}\left(\Delta_i\right)^2\right).
\]
Therefore, by Lemma~\ref{BirgeMassart}, letting $\gamma = n \theta (1 - \theta) \sum_{i \in \mathcal{S}} \left(\Delta_i\right)^2$
\[
\mathbb{P}\left(\mbox{Diff} + \beta + |\mathcal{S}|\alpha \geq |\mathcal{S}| + \gamma - 2\sqrt{\left(|\mathcal{S}| + 2 \gamma\right)y} \right) > 1 - \exp(-y).
\]
Note that if $\mbox{Diff} > 0$ then a changepoint will be detected. Therefore we require that
\[
\gamma \geq 4 y + \beta + |\mathcal{S}|(\alpha - 1) + \sqrt{4y\left((2\alpha - 1)|\mathcal{S}| + 2 \beta + 4 y\right)}.
\]
We may set $y = \delta \log a$, for $a = \max\{n, d\}$, to give that $\mathbb{P}\left(\mbox{Diff} > 0 \right) > 1 - a^{-\delta}$, providing that
\[
\sum_{i \in \mathcal{S}} \left(\Delta_i\right)^2 \geq \frac{4 \delta \log a  + \beta + |\mathcal{S}|(\alpha - 1) + 2\sqrt{ \delta \log a \left((2 \alpha - 1) |\mathcal{S}| + 2 \beta + 4 \delta \log a \right)}}{n \theta \left(1 - \theta\right)},
\]
as required. \QEDB

\textbf{Proof of Lemma~\ref{densepower}}: When comparing a fit at the true location $\tau = \theta n$ under a total penalty of $K$ to the null fit, the difference in cost (in favour of the the non-null fit) is distributed as a non-central chi-squared distribution with $d$ degrees of freedom and non-centrality parameter $n \theta \left(1 - \theta\right) \sum_{i=1}^d \left(\Delta_i\right)^2$. By Lemma~\ref{BirgeMassart} and the definition of $K$, we therefore see that setting $\nu - 2 \sqrt{y}\sqrt{d + 2 \nu} \geq 2 \sqrt{dx} + 2x$ for $\nu = n \theta \left(1 - \theta\right) \sum_{i=1}^n \left(\Delta_i\right)^2$ gives that $\mathbb{P}\left(\chi_d^2\left(\nu\right) > K\right) \geq 1 - \exp\left(-y\right)$.

Resolving the inequality $\nu - 2 \sqrt{y}\sqrt{d + 2 \nu} \geq 2 \sqrt{d x} + 2x$ gives that

\begin{equation}\label{nucondition}
\nu \geq 4y + 2x + 2\sqrt{xd} + 2 \sqrt{y\left(4y + \left(\sqrt{4x} + \sqrt{d}\right)^2\right)};
\end{equation}

as $x = \beta/2$, and setting $y = \delta \log a$, (\ref{nucondition}) becomes
\[
\sum_{i=1}^d \left(\Delta_i\right)^2 \geq \frac{4 \delta \log a + K - d + 2 \sqrt{\delta \log a \left(4 \delta \log a + 2 K - d\right)}}{n \theta \left(1 - \theta\right)},
\]
as required. \QEDB

With these lemmas, we are now in a position to prove the results of the main article.

\section{Proofs of Main Results}\label{sec:proofs}

In this section, we combine the preliminary results of Section~\ref{sec:prelimlemmas} to give proofs of the results stated in the main article.

\textbf{Proof of Theorem~\ref{allparameterchoices}}: From Lemma~\ref{typeIsparse}, letting $g\left(n,d\right) = d$ and $C=\sqrt{J + \varrho}$, some $\varrho > 0$, gives that $\alpha = 2 \log d$ and $\sqrt{\beta} = \sqrt{2d \frac{\Gamma\left(\frac{1}{2}, \log d\right)}{\Gamma\left(\frac{1}{2}\right)}} + \sqrt{J + \varrho} \sqrt{\log n}$, so that in the sparse setting
\[
\mathbb{P}\left(\max_t S_{1,t} > 0\right) \leq n^{1 - \frac{J}{2} -\varrho/2} \exp\left(\frac{(1+\vartheta)^2}{4 \vartheta}\frac{1}{\sqrt{1 + 2 \log d}} \right),
\]
where here $\frac{\gamma\left(\frac{1}{2},\frac{\alpha}{2}\right)}{\Gamma\left(\frac{1}{2}\right)} \geq \vartheta > 0$.

As we have $\frac{\Gamma\left(s,x\right)}{\Gamma\left(s\right)} \leq e^{-x} \left(1 + \frac{x}{s}\right)^{s-1}$ for $0 < s < 1$, we have that
\[
\vartheta \leq 1 - \frac{1}{d\left(1 + \log d\right)^{\frac{1}{2}}}
\]
so that, for example, by taking $d = 2$, we may bound $\exp\left(\frac{(1 + \vartheta)^2}{4\vartheta} \frac{1}{\sqrt{1 + 2 \log d}}\right)$ above by an absolute constant $\forall d \geq 2$. For $\max_t S_{2,t}$, we use Lemma~\ref{typeIdense}, and the result for SUBSET in both settings follows as $S_t = \max_t \left\{S_{1,t}, S_{2,t}\right\}$. \QEDB

\textbf{Proof of Theorem~\ref{sufficiencycondition}}: In the sparse setting, we may directly apply Lemma~\ref{sparsepower}, while in the dense setting, we may directly apply Lemma~\ref{densepower}. Note that the condition in Lemma~\ref{sparsepower} resolves to give the required statement by setting $K_{\mathcal{S}} = \beta +|\mathcal{S}|\alpha$. \QEDB

\section{SUBSET Pseudo-Code, Post-Processing and Computational Discussion}\label{sec:postproc}

Algorithm 1 gives the pseudo-code of the SUBSET procedure without the post-processing step.

\footnotesize
\begin{figure}
\begin{algorithm}[H]
\small
 \caption{SUBSET (without post-processing).}
 \KwData{A multivariate dataset, $\left(y_{i,j}\right)_{i=1, \ldots, d, j =1, \ldots, n}$; variate penalty, $\alpha$; changepoint penalty function, $\beta$; threshold penalty function, $K$; segment cost function, $\mathcal{C}\left(.\right)$; an interval number, $M$; a sort function with respect to vector $\textbf{v}$, $\rho_\textbf{v}\left(.\right)$.}
 \KwResult{An estimated set of changepoints $\hat{\tau}_1, \ldots, \hat{\tau}_{\hat{m}}$ and corresponding estimated affected sets $\hat{\mathcal{S}}_{1}, \ldots, \hat{\mathcal{S}}_{{\hat{m}}}$.}
 \textbf{Step 0}: Set $l_0=1$, $u_0=n$, $\boldsymbol{\hat{\tau}} = NULL$, $\boldsymbol{\hat{\mathcal{S}}} = NULL$. \For{$j \in \left\{1, \ldots, M\right\}$}{
   $r_j \sim U\left\{1, \ldots, n\right\}$, $s_j \sim U\left\{1, \ldots, n\right\}$, $(l_j, u_j) = \left(\min\{r_j, s_j\}, \max\{r_j, s_j\} \right)$
  }
 \textbf{Step 1}: \For {$j \in \left\{0, \ldots, M\right\}$}{
	\uIf{$u_j - l_j > 1$, $l_j \geq l_0$, $u_j \leq u_0$}{
		\For{$t \in \left\{l_j, \ldots, u_j\right\}$}{
		 $S_{1,t} = \sum_{i=1}^d \max\left\{\mathcal{C}\left(y_{i,l_j:u_j}\right) - \mathcal{C}\left(y_{i,l_j:t}\right) - \mathcal{C}\left(y_{i,(t+1):u_j}\right) - \alpha, 0\right\}$ \\
		 $S_{2,t} = \sum_{i=1}^d \left\{ \mathcal{C}\left(y_{i,l_j:u_j}\right) - \mathcal{C}\left(y_{i,l_j:t}\right) - \mathcal{C}\left(y_{i,(t+1):u_j}\right) \right\}$ \\
		 $S_t = \max\left\{S_{1,t} - \beta, S_{2,t} - K \right\}$
		 		 
		}
		\uIf{$\max_t S_t > 0$}{ $q^{j} = \arg \max S_t$, $T_{q^j} = \max S_t$\\
			\uIf{$S_{q^j} = S_{1,q^j} - \beta$}{
				$\mathcal{T}_{q^j} = \left\{i : \mathcal{C}\left(y_{i,l_j:u_j}\right) - \mathcal{C}\left(y_{i,l_j:q^j}\right) - \mathcal{C}\left(y_{i,(q^j +1):u_j} \right) - \alpha > 0 \right\}$
			}
			\Else{
				$\mathcal{T}_{q^j} = \left\{1, \ldots, d\right\}$
			}
		}
		\Else{
			$\left(q^j, T_{q^j}, \mathcal{T}_{q^j}\right) = \left(NULL, 0, \emptyset\right)$
		}
	}
	\Else{
		$\left(q^j, T_q^j, \mathcal{T}_{{q}^{j}}\right) = \left(NULL, 0, \emptyset\right)$
	}
}
 \textbf{Step 2}: Set $\textbf{q} = \left(q^0, q^1, \ldots, q^{M} \right)$ \\
	\uIf{$||\textbf{q}||_0\geq1$ }{
		$\gamma = \arg \max _{j \in \left\{1, \ldots, M+1\right\}} T_{q^j}$, $\eta = q^{\gamma}$, $\mathcal{U} = \mathcal{T}_{\eta}$, $\boldsymbol{\hat{\tau}} = \left(\boldsymbol{\hat{\tau}}, \eta \right)$, $\boldsymbol{\hat{\mathcal{S}}} = \left(\boldsymbol{\hat{\mathcal{S}}}, \mathcal{U}\right)$ \\
		$\boldsymbol{\hat{\tau}} = \rho_{\boldsymbol{\hat{\tau}}} \left(\boldsymbol{\hat{\tau}}\right)$, $\boldsymbol{\hat{\mathcal{S}}} = \rho_{\boldsymbol{\hat{\tau}}} \left(\boldsymbol{\hat{\mathcal{S}}}\right)$ \\
		Return to Step 1 with $l_0 = l_0$ and $u_0 = \eta$, and return to Step 1 with $l_0 = \eta + 1$ and $u_0 = u_0$. \\
	}
	\Else{
		%$\eta = NULL$, $\mathcal{U} = \emptyset$
		If there are no further `active' $(l_0, u_0)$ intervals, then return $\boldsymbol{\hat{\tau}}, \boldsymbol{\hat{\mathcal{S}}}$.
	}

\vspace{10pt}
\end{algorithm}
\end{figure}
\normalsize

As discussed in Section~\ref{ssec:SUBSET} of the main article, a post-processing step is required in the SUBSET procedure to ensure that masking between different changepoints present in the data do not cause misspecification in the estimates of the affected sets at each changepoint. We detail this post-processing procedure in Algorithm 2.

\begin{algorithm}[H]
 \caption{Post-processing step for the SUBSET procedure.}
 \KwData{A multivariate dataset, $\textbf{y}_{1:n}$; an $\alpha$ and $\mathcal{C}\left(.\right)$ as for Algorithm 1 of the main article; a set of candidates returned by Algorithm 1 of the main article, $0 = \xi_0 < \xi_1 < \ldots < \xi_q < \xi_{q+1} = n$.}
 \KwResult{An estimated set of changepoints $\hat{\tau}_1, \ldots, \hat{\tau}_{\hat{m}}$ and corresponding estimated affected sets $\hat{\mathcal{S}}_{1}, \ldots, \hat{\mathcal{S}}_{{\hat{m}}}$.}
 
 \textbf{Step 0}: Set $\hat{\mathcal{S}}_{1} = \ldots = \hat{\mathcal{S}}_{q} = \emptyset$, $\hat{\tau} = NULL$;
 
 \For{$i \in \left\{1, \ldots, d \right\}$}{
 	$F =  \left(- \alpha, 0, \ldots, 0\right)$; \\
 	\For{$j \in \left\{1, \ldots, q + 1 \right\}$}{
 		$F[j+1] = \min\limits_{1 \leq k \leq j} \left[F[k] + \mathcal{C}\left(y_{i,\xi_{k-1}:\xi_{j}}\right) + \alpha \right]$; \\
 		$r = \arg\min\limits_{1 \leq k \leq j} \left[F[k] + \mathcal{C}\left(y_{i,\xi_{k-1}:\xi_{j}}\right) + \alpha \right]$; \\
 		$\hat{\mathcal{S}}_{{r-1}} = (\hat{\mathcal{S}}_{{r-1}}, \left\{i\right\})$ \\
 	}
 }

\For{$j \in \{1, \ldots, q\}$}{
	\If{$\hat{\mathcal{S}}_{j} \neq \emptyset$}{
		$\hat{\tau}  = (\hat{\tau}, \xi_j)$
	}
}

\vspace{10pt}
\end{algorithm}

Note that this procedure, which closely parallels the Optimal Partitioning of \cite{Jacksonetal05}, has complexity of $\mathcal{O}\left(q^2 d\right)$, where $q$ is the number of candidate changepoint locations returned by SUBSET. Indeed, employing a pruning step as per the PELT procedure of \cite{KillickEckleyFearnhead12} results in an expected cost of $\mathcal{O}\left(qd\right)$. As shown in \cite{TickleEckleyFearnhead18}, this can be improved further to a worst-case cost of $\mathcal{O}\left(qd\right)$ using parallelisation. Therefore, the worst-case computational complexity of the post-processing step is $\mathcal{O}\left(nd\right)$.

Given that the SUBSET procedure employs a hybrid of a Wild Binary Segmentation approach, simulating $M$ intervals at each stage, the worst-case computational cost of SUBSET is not dominated by the post-processing step, and is $\mathcal{O}\left(M d n \log n \right)$.

\section{Additional Material on the Analysis of the Global Terrorism Database}\label{sec:realdataextra}

\begin{figure}[h!]
\begin{minipage}[t]{\textwidth}
\vspace{-10pt}
\begin{center}
\includegraphics[width=0.98\linewidth,keepaspectratio=TRUE]{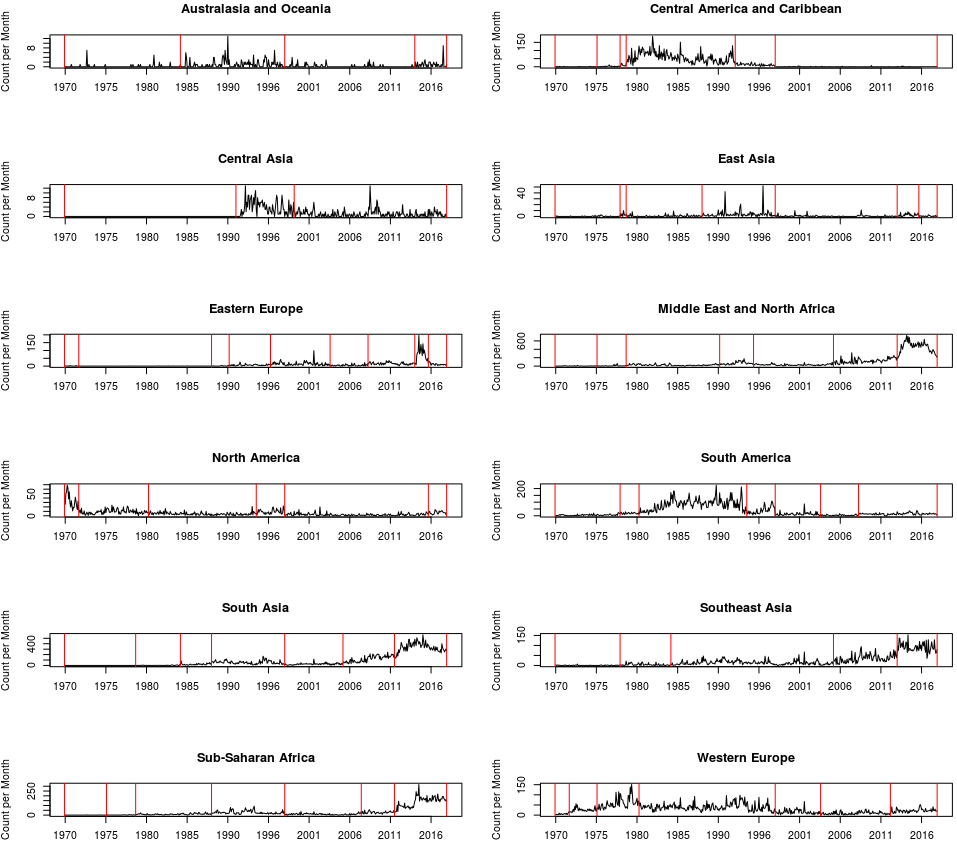}
\end{center}
\vspace{-20pt}
\caption{Incident count for each region between 1970 and 2017 with changes found by the SUBSET method overlaid as red vertical lines.}
\label{terrorchanges}
\end{minipage}
\end{figure}

\begin{figure}[h!]
\begin{minipage}[t]{\textwidth}
\vspace{-10pt}
\begin{center}
\includegraphics[width=0.98\linewidth,keepaspectratio=TRUE]{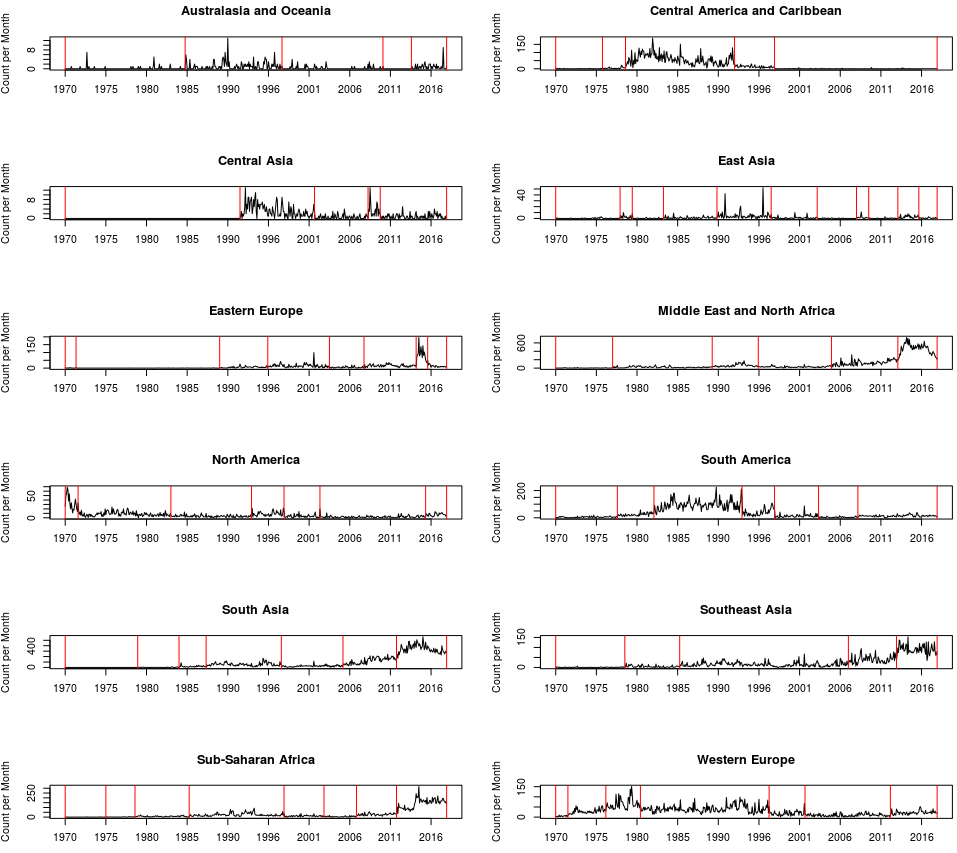}
\end{center}
\vspace{-20pt}
\caption{Incident count for each region between 1970 and 2017 with changes for individual series found by a univariate method overlaid as red vertical lines.}
\label{singleterrorchanges}
\end{minipage}
\end{figure}

After running SUBSET through the time series, the estimated changepoints and the corresponding affected sets of regions were computed. These are shown in Table~\ref{twopenincidents}. For an alternative visualisation, with the estimated changes for each region superimposed over the raw count data, see Figure~\ref{terrorchanges}. By comparison, Figure~\ref{singleterrorchanges} shows the results of applying a univariate method (in this case the minimisation of the same penalised univariate negative binomial cost function using dynamic programming) to each series individually.

\begin{table}
\begin{tabular}{ |p{1.6cm}|p{14.0cm}|}
 \hline
 \multicolumn{1}{|c|}{Dates} & \multicolumn{1}{|c|}{Regions} \\
 \hline
 Sep 1971 & E.Eu, N.Am, W.Eu \\
 Feb 1975 & C.Am \& C, M.E. \& N.Af, SS.Af, W.Eu \\
 Dec 1977 & C.Am \& C, E.As, S.Am, SE.As \\
 Sep 1978 & C.Am \& C, E.As, M.E. \& N.Af,  S.As, SS.Af\\
 Apr 1980 & N.Am, S.Am, W.Eu \\
 Mar 1984 & Au \& Oc, S.As, SE.As \\
 Jan 1988 & E.As, E.Eu, S.As, SS.Af \\
 Mar 1990 & E.Eu, M.E. \& N.Af \\
 Jan 1991 & C.As \\
 Feb 1992 & C.Am \& C \\
 Jul 1994 & N.Am, S.Am \\
 May 1995 & M.E. \& N.Af \\
 Apr 1996 & E.Eu \\
 Jan 1998 & Au \& Oc, C.Am \& C, E.As, N.Am, S.Am, S.As, SS.Af, W.Eu\\
 Mar 1999 & C.As \\
 Aug 2003 & E.Eu, S.Am, W.Eu \\
 Mar 2005 & M.E. \& N.Af, S.As, SE.As \\
 Jun 2007 & SS.Af \\
 Apr 2008 & E.Eu, S.Am \\
 Jul 2011 & S.As, SS.Af  \\
 Mar 2012 & W.Eu \\
 Jan 2013 & E.As, M.E. \& N.Af, SE.As \\
 Jan 2014 & Au \& Oc, E.Eu \\
 Sep 2015 & E.As, E.Eu, N.Am \\
 \hline
\end{tabular}
\caption{Changepoints found within the count data of terrorist incidents per month using the SUBSET procedure. The regions column corresponds to those areas which are said to be affected by the corresponding changepoint.}
\label{twopenincidents}
\end{table}

Several salient features of the dataset are revealed by this analysis. Firstly, we note that there are many similarities between the changes found by the univariate method and SUBSET: for several of the series (for example, Western Europe), the same number of changepoints are found, with broadly the same change locations. However, in general, we see that SUBSET is more parsimonious. In addition, by its nature, changepoints which occur in different series at the same time are more readily identified by SUBSET. For example, the most dense changepoint (following post-processing) located using SUBSET is that of January 1998, a month corresponding to a change in the data collection methods for the GTD for the ``GTD2'' phase.

Other points of interest found by SUBSET include several ``staggered'' changepoints at the end of the 1980s and the beginning of the 1990s, which appear to correspond to the end of the Cold War. These findings are intuitive in the sense that many countries formerly in the Soviet sphere began publishing incidents in media outlets available to the investigators compiling the GTD. More recent changepoints seem to align to significant events in, for example, the Arab Spring uprising and the conflict in Ukraine.

\newpage

\subsection{SUBSET and the Global Terrorism Database}\label{ssec:subsetgtd}

We now seek to verify the robustness of our findings after applying SUBSET to the Global Terrorism Database. The most fundamental assumption made by the SUBSET procedure in undertaking this analysis is that each series corresponding to a given global region has broadly stationary residuals, with no or low correlation with other regions, up to the location of the changepoints. 

\begin{figure}[h!]
\begin{minipage}[t]{\textwidth}
\vspace{-10pt}
\begin{center}
\includegraphics[width=0.98\linewidth,keepaspectratio=TRUE]{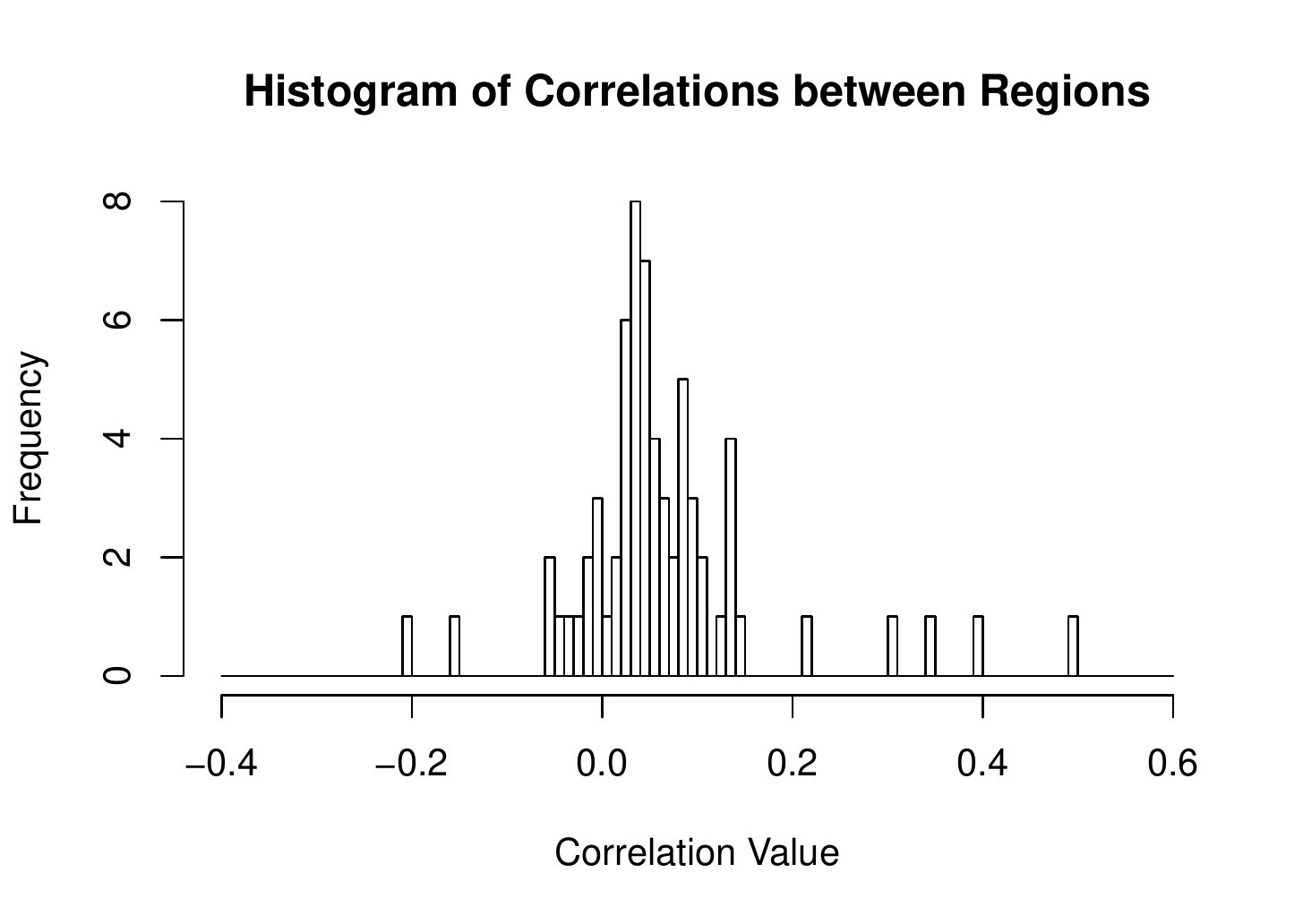}
\end{center}
\vspace{-20pt}
\caption{Histogram of correlations between residuals in different regions, corrected for changepoints located in each series in each case.}
\label{fig:corrhist}
\end{minipage}
\end{figure}

Using the changepoint model returned to us by SUBSET in Table~\ref{twopenincidents}, we calculate the Pearson residuals for each geographical region.  For each pair or regions we then calculated the correlation between these residuals, with the results shown in Figure~\ref{fig:corrhist}. The mean correlation between a pair of regions is 0.063 to three decimal places.

%Using the changepoint model returned to us by SUBSET in Table~\ref{twopenincidents}, we compute the (Pearson) residuals of each region between each of the estimated changepoint locations. The results are shown in the histogram in Figure~\ref{fig:corrhist}, which indicates that no two series have a greater correlation than $0.5$.

\end{document}